\documentclass[journal]{IEEEtran}
\usepackage{caption} 
\usepackage{soul}
\usepackage{lipsum}
\usepackage{indentfirst}
\usepackage{booktabs}
\usepackage{enumerate}
\usepackage{tabu}
\usepackage{tabularx}
\usepackage{multirow}
\usepackage{environ}         
\usepackage{etoolbox}        
\usepackage[dvipsnames]{xcolor}

\usepackage{cite}

\usepackage{graphicx}
\usepackage{wrapfig}
\usepackage[labelformat=simple]{subcaption}

\usepackage{float}
\usepackage{eurosym}
\usepackage{gensymb}
\usepackage{textpos}

\usepackage{amsfonts, eqnarray, amsmath, amssymb}
\usepackage{dsfont}
\usepackage{amsthm}
\usepackage{mathtools}
\usepackage{xfrac}
\usepackage{breqn}
\usepackage{bm}

\usepackage[pscoord]{eso-pic}
\newcommand{\placetextbox}[3]{
  \setbox0=\hbox{#3}
  \AddToShipoutPictureFG*{
    \put(\LenToUnit{#1\paperwidth},\LenToUnit{#2\paperheight}){\vtop{{\null}\makebox[0pt][c]{#3}}}%
  }%
}%

\usepackage{algorithm}
\usepackage{algpseudocode}
\usepackage{etoolbox}
\algnewcommand{\LeftComment}[1]{\Statex \(\triangleright\) #1}

\makeatletter
\newcommand*{\algrule}[1][\algorithmicindent]{%
  \hspace*{.2em}
  \vrule 
  \hspace*{\dimexpr#1-.2em-.4pt}%
}

\newcommand{\StatePar}[1]{%
  \State\parbox[t]{\dimexpr\linewidth-\ALG@thistlm}{\strut #1\strut}%
}
\renewcommand{\ALG@beginalgorithmic}{\offinterlineskip}

\newcount\ALG@printindent@tempcnta
\def\ALG@printindent{%
  \ifnum \theALG@nested > 0
    \ifx\ALG@text\ALG@x@notext
    \else
      \unskip
      \ALG@printindent@tempcnta=1
      \loop
        \algrule[\csname ALG@ind@\the\ALG@printindent@tempcnta\endcsname]%
        \advance \ALG@printindent@tempcnta 1
        \ifnum \ALG@printindent@tempcnta<\numexpr\theALG@nested+1\relax
      \repeat
        \fi
    \fi
}
\patchcmd{\ALG@doentity}{\noindent\hskip\ALG@tlm}{\ALG@printindent}{}{\errmessage{failed to patch}}
\makeatother

\algrenewcommand\algorithmicend{\strut\textbf{end}}
\algrenewcommand\algorithmicdo{\strut\textbf{do}}
\algrenewcommand\algorithmicwhile{\strut\textbf{while}}
\algrenewcommand\algorithmicfor{\strut\textbf{for}}
\algrenewcommand\algorithmicforall{\strut\textbf{for all}}
\algrenewcommand\algorithmicloop{\strut\textbf{loop}}
\algrenewcommand\algorithmicrepeat{\strut\textbf{repeat}}
\algrenewcommand\algorithmicuntil{\strut\textbf{until}}
\algrenewcommand\algorithmicprocedure{\strut\textbf{procedure}}
\algrenewcommand\algorithmicfunction{\strut\textbf{function}}
\algrenewcommand\algorithmicif{\strut\textbf{if}}
\algrenewcommand\algorithmicthen{\strut\textbf{then}}
\algrenewcommand\algorithmicelse{\strut\textbf{else}}

\algrenewcommand\algorithmicrequire{\strut\textbf{Input:}}
\algrenewcommand\algorithmicensure{\strut\textbf{Output:}}

\let\oldState\State
\renewcommand{\State}{\oldState\strut}

\usepackage{array}

\usepackage{url}

\NewEnviron{resizealign}{\sbox0{
    $\begin{matrix}\displaystyle\BODY\end{matrix}$}%
  \sbox1{$(\theequation)$}%
  \sbox2{\parbox{\dimexpr \wd0 + 2\wd1}%
    {\begin{align}\BODY\end{align}}}
  \noindent\resizebox{\columnwidth}{!}{\usebox2}%
}

\begin{document}
%
\title{User Head Movement-Predictive XR in Immersive H2M~Collaborations~over~Future~Enterprise~Networks}
%
\author{Sourav Mondal,~\IEEEmembership{Member,~IEEE},
        and Elaine Wong,~\IEEEmembership{Senior Member,~IEEE, Fellow,~OSA} %
\thanks{S. Mondal and E. Wong are with the Faculty of Engineering and Information Technology, Department of Electrical and Electronic Engineering, University of Melbourne, Parkville, VIC 3010, Australia (email: \{sourav.mondal,ewon\}@unimelb.edu.au).}}

\markboth{IEEE INTERNET OF THINGS JOURNAL,~Vol.~XX, No.~X, XXX~2025}%
{Shell \MakeLowercase{\textit{et al.}}: Bare Demo of IEEEtran.cls for IEEE Journals}


\placetextbox{0.5}{0.04}{This article is accepted for publication in IEEE Internet of Things Journal. Copyright @ IEEE 2025.}%

\maketitle

\begin{abstract} 
The evolution towards future generation of mobile systems and fixed wireless networks is primarily driven by the urgency to support high-bandwidth and low-latency services across various vertical sectors. This endeavor is fueled by smartphones as well as technologies like industrial internet of things, extended reality (XR), and human-to-machine (H2M) collaborations for fostering industrial and social revolutions like Industry 4.0/5.0 and Society 5.0. \textcolor{black}{To ensure an ideal immersive experience and avoid cyber-sickness for users in all the aforementioned usage scenarios, it is typically challenging to synchronize XR content from a remote machine to a human collaborator according to their head movements across a large geographic span in real-time over communication networks. Thus, we propose a novel H2M collaboration scheme where the human's head movements are predicted ahead with highly accurate models like bidirectional long short-term memory networks to orient the machine's camera in advance. We validate that XR frame size varies in accordance with the human's head movements and predict the corresponding bandwidth requirements from the machine's camera to propose a human-machine coordinated dynamic bandwidth allocation (HMC-DBA) scheme. Through extensive simulations, we show that end-to-end latency and jitter requirements of XR frames are satisfied with much lower bandwidth consumption over enterprise networks like Fiber-To-The-Room-Business.} Furthermore, we show that better efficiency in network resource utilization is achieved by employing our proposed HMC-DBA over state-of-the-art schemes.
\end{abstract}

\begin{IEEEkeywords}
BiLSTM, FTTR-Business, immersive human-to-machine collaborations, time-series, XR communications. 
\end{IEEEkeywords}

\IEEEpeerreviewmaketitle

\section{Introduction} \label{sec1}
\IEEEPARstart{A}{longside} the continuous evolution of smartphones, mobile industries across the globe are strongly influencing the development and deployment of future mobile and wireless broadband networks to support vertical sectors like smart industries and manufacturing, automotive, e-healthcare, and agriculture, to name a few \cite{5g_vert}. In the industry vertical revolution, Industry 4.0 focused on fully automated smart manufacturing with the aid of cyber-physical systems, big data, cloud computing, artificial intelligence (AI), and industrial internet-of-things \cite{xr_5G}. However, Industry 5.0 reinvented the role of human intelligence for mass customization and personalization through human-to-machine (H2M) collaborations \cite{h2m,ind5.0,ind_robo}. In this endeavor, some of the primary technologies involved are immersive XR encompassing augmented, mixed, and virtual reality and multi-sensory communications \cite{glad,martin_robo,imrsv}. The core requirement to ensure \emph{ideal immersive quality of experience (QoE) for XR in H2M collaborations} is to ensure that very-high quality XR video frames are transmitted while satisfying the end-to-end latency and jitter requirements.\par
\begin{figure}[!t]
\centering
\includegraphics[width=0.99\columnwidth]{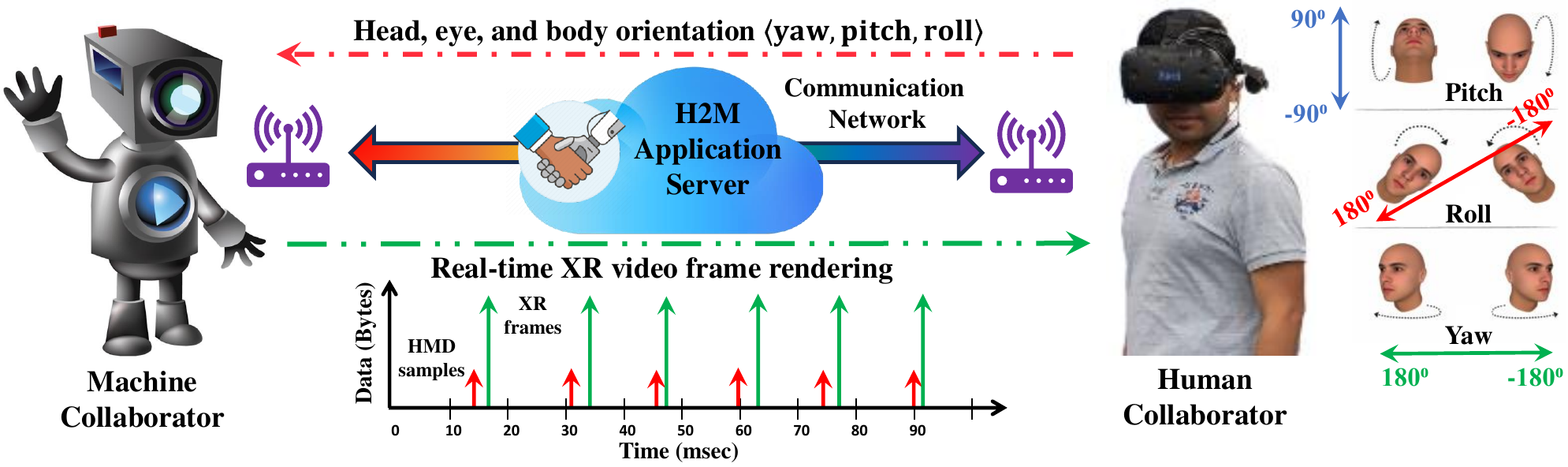}
\caption{A schematic showing H2M collaboration over a network.}
\label{h2m}
\end{figure}

\begin{table}[!t]
\centering
\caption{QoE Requirements for XR in H2M collaborations}
\label{table1}
\resizebox{\columnwidth}{!}{%
\begin{tabular}{lccc}
\toprule
                                                                                           & \textbf{Fair QoE}                                         & \textbf{Comfortable QoE}                                      & \textbf{Ideal QoE}                                                                   \\ \toprule
\textbf{\begin{tabular}[c]{@{}l@{}}Strongly-interactive\\ content resolution\end{tabular}} & \begin{tabular}[c]{@{}c@{}}2K\\ {[}60 fps{]}\end{tabular} & \begin{tabular}[c]{@{}c@{}}4K\\ {[}60, 90 fps{]}\end{tabular} & \begin{tabular}[c]{@{}c@{}}8K to 16K\\ {[}60, 90, 120 fps{]}\end{tabular}            \\ \hline
\textbf{Datarate}                                                                          & $\geq 40$ Mbps                                            & $\geq 90$ Mbps                                                & \begin{tabular}[c]{@{}c@{}}$\geq 360$ Mbps (8K)\\ $\geq 440$ Mbps (16K)\end{tabular} \\ \hline
\textbf{Bandwidth}                                                                         & $\geq 80$ Mbps                                            & $\geq 260$ Mbps                                               & \begin{tabular}[c]{@{}c@{}}$\geq 1$ Gbps (8K)\\ $\geq 1.5$ Gbps (16K)\end{tabular}   \\ \hline
\textbf{Inter-frame latency}                                                                & $\leq 20$ msec                                            & $\leq 15$ msec                                                & $\leq 8$ msec                                                                        \\ 
\hline
\end{tabular}
}
\end{table}
\setlength{\textfloatsep}{3pt}
\vspace{-0.5\baselineskip}
In augmented and mixed reality systems, the human's head-mounted device (HMD) sends \emph{sensing information} like head and eye tracking and real-time videos of real environments to the application server. The server augments some virtual contents, and the video is rendered back to the user. In virtual reality systems, the human's HMD sends only sensing information to the application server to render corresponding pre-created video content \cite{xr_std}. On the other hand, in H2M collaborations, sensing information from the human's HMD is sent to the application server that translates it into corresponding camera orientation commands for the machine, as shown in Fig. \ref{h2m}. Accordingly, the machine orients its camera, and subsequently, real-time video frames from the machine's camera are transmitted to the human's HMD.\par
\vspace{-0.5\baselineskip}
\textcolor{black}{Nonetheless, in practice, a time lag between the human's head and the robot's camera movements may occur due to mismatch between their rotational speeds or data transmission latency across a large geographic span. If this time lag becomes consistently $\geq 20$ msec, then human brains detect it as a de-synchronization between head movements and the visual scene, thus causing cyber-sickness \cite{vr_bk}. Therefore, it is critically important to synchronize the movements of the human's head and the machine's camera within this threshold for an ideal immersive experience and to ensure that the humans do not suffer from cyber-sickness after prolonged exposure to H2M collaborations.} As vision is the strongest among all human senses, synchronized visual information can greatly compensate any discomforts arising from minor jitter in haptics and multi-sensory information.\par
\vspace{-0.5\baselineskip}
It is important to note that high-quality XR video rendering for H2M collaborations not only demands high datarate but also low-latency with low-jitter between consecutive frames \cite{itu_fttr}, as summarized in Table \ref{table1}. For example, 4K video with 60 frames/sec requires a datarate of nearly 90 Mbps with 10-30 msec end-to-end packet latency budget. Note that even though the peak datarate of 5G mobile systems is specified to be 20 Gbps, the practically achievable datarate is approximately 0.1-2.0 Gbps. Thus, efficient management of scarce network resources in next-generation access networks is critically important to satisfy desired QoE. In this regard, the authors of \cite{ml_ewon} extensively explored some of the primary challenges and solutions for network resource allocation in next-generation networks. Application layer and network layer collaborative approaches like network-aware adaptive XR video streaming \cite{xr_5g1} and XR QoE-aware scheduling \cite{xr_5g2} have been proposed as promising solutions. Alongside access networks, core and edge servers also play an important role by performing field-of-view (FOV) projection, spatial warping, and temporal warping that facilitates XR in H2M collaborations. The authors of \cite{xr_survey} provided a detailed survey on recent XR solutions like FOV-dependent streaming, split rendering, and context-aware XR computing, to name a few.\par
\begin{figure*}[!t]
\centering
\includegraphics[width=0.97\textwidth,keepaspectratio]{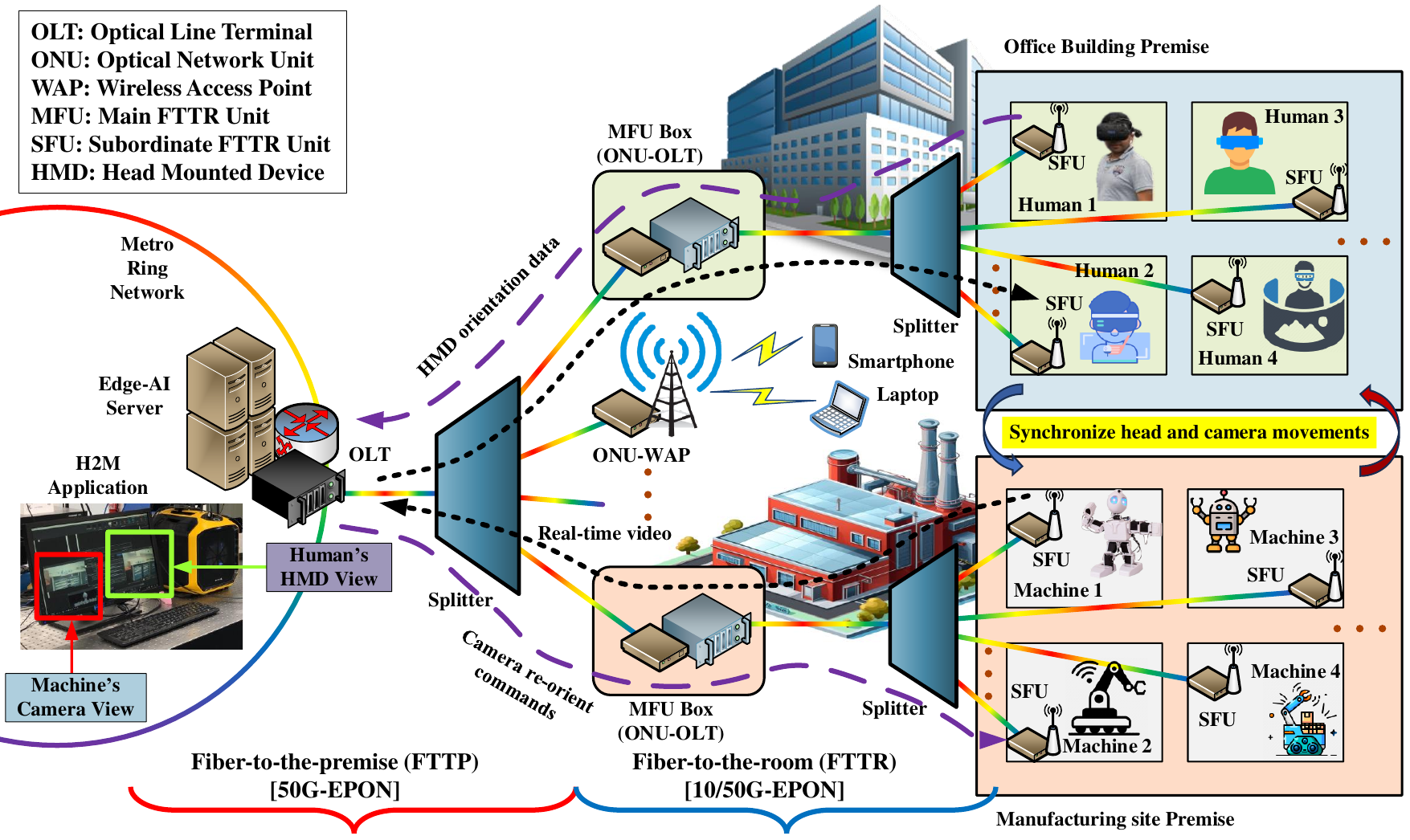}
\caption{Immersive H2M application over FTTR-Business networks. Prediction of the human's head movement allows the remote machine's camera to be pre-oriented, thus enabling transmission of synchronized XR contents to the human's HMD.}
\label{hcs}
\end{figure*}

\setlength{\textfloatsep}{5pt}
%
Recent market predictions indicate that the XR market is expected to grow at an annual rate of 8.97\% over 2024-2029 to reach a projected market volume of \$62 billion \cite{xr_stat}. To satisfy network resource demand of this large number of devices while meeting the QoE of their respective services, the natural evolution of high-capacity Fiber-To-The-x technology into Fiber-To-The-Room (FTTR) for premium home broadband and FTTR-Business for industrial applications are being considered as potential solutions \cite{fttr_jocn}, as shown in Fig. \ref{hcs}. The first segment of the FTTR architecture is an external time-division multiplexed passive optical network (TDM-PON) supporting 50 Gbps datarate, and the second segment is an in-premise TDM-PON supporting 10/50 Gbps datarate. The optical line terminal (OLT) of the first segment connects multiple main FTTR units (MFUs), a special type optical network unit (ONU). Each MFU, in turn, connects multiple subordinate FTTR units (SFUs) supporting WiFi 6/6E, WiFi 7, or 5G mobile wireless access points (WAPs). To the best of our knowledge, \emph{the end-to-end latency of XR in H2M collaborations over FTTR networks is yet to be investigated.}\par
In H2M collaborations, the human's head orientations are transmitted either in Euler angles (i.e., pitch, yaw, roll) or in quaternion format to the H2M application server over a communication network, as shown in Fig.~\ref{h2m}. In turn, this information is converted into command instructions and transmitted for orienting the machine's camera to a similar position. Usually, motors and gears are involved to perform the rotational tasks. After orienting the camera to the intended direction, real-time XR content is transmitted for the H2M application to format and render at the human's HMD. In practice, the camera rotation is done by servo motors operating at a certain operating speed (e.g., 15 degree/sec), and a sequence of positive or negative voltage pulses is required to rotate in the right or left direction. \textcolor{black}{A set of the typical actions during a H2M collaboration like picking up some object, placing an object on a rack, tightening a screw with a screwdriver, to name a few, are listed in \cite{ind_act}. Moreover, the speed of human head rotations can vary ($\geq 180$ degree/sec) depending on the activity \cite{head_speed}. Therefore, if the speed of motor rotation is much slower than the human's head movement, then the human may experience cyber-sickness. In this work, we address this significant issue by predicting the human's future head orientations and consequently sending camera rotation commands to the machine in advance. Note that the \emph{prediction horizon} can be chosen based on the rotational speed gap between human's head and machine's motors (can be measured at the initial calibration stage) to make the framework generic and task-agnostic.}\par
\textcolor{black}{We observe from our analysis that one of the primary factor influencing XR frame size is the angular shift of human's head as it determines the correlation between consecutive frames (please refer to Sec. \ref{sec3.2}). Hence, during sudden and rapid head movements, a large angular shift between consecutive XR frames reduces inter-frame correlation, resulting in increased frame sizes due to the need to transmit new FOV tiles or pixels \cite{xr_quality}. At such instances, the demand for network bandwidth increases abruptly to transmit the larger XR frames. If the required bandwidth is not immediately available, some packets may be delayed, potentially leading to XR frame decoding failures and degradation in Quality of Experience (QoE).} \textcolor{black}{To address this challenge, we propose a novel \emph{human-machine coordinated dynamic bandwidth allocation (HMC-DBA) scheme} where each human's specific future head movements from the H2M application are conveyed to the main OLT for bandwidth prediction and resource allocation.} As the camera rotation commands are sent to the machine in advance, the camera starts to rotate at a steady speed before the human's actual head movement and the resource demand increases gradually over a longer duration. In doing so, the probability of delayed packet transmission is reduced and a satisfactory QoE can be maintained. Moreover, exploiting the pseudo-periodic characteristics of the XR and HMD orientations traffic \cite{xr_tmp}, we pre-allocate resources to further reduce transmission latency and jitter. Overall, our primary contributions in this paper are summarized below.
\vspace{-0.3\baselineskip}
\begin{itemize}
    \item \textcolor{black}{We perform an experimental study of the H2M collaboration to understand the statistical characteristics of HMD and XR traffic. We derive mathematical models to estimate XR frame size from human's head movements and validate it with experimental data.}
    \item As the statistical characteristics of HMD traffic showed the prediction challenges, we employ multiple methods like persistence, moving-average, auto-regressive integrated moving average (ARIMA), and bi-directional long short-term memory network (BiLSTM) methods, observing a normalized RMSE around 10\% even for 90 msec ahead prediction with head movement at $180\degree$/sec.  
    \item \textcolor{black}{We design an end-to-end framework over FTTR-Business network architecture for practical deployment of XR in H2M collaborations by adopting our proposed HMC-DBA scheme. We perform discrete-event network simulations to show that ideal immersive experience is ensured by this framework even under network conditions where state-of-the-art resource allocation schemes fail.}
\end{itemize}
\setlength{\textfloatsep}{1pt}
\par The rest of this paper is organized as follows. Sec.~\ref{sec2} reviews some recent related works. Sec.~\ref{sec3} briefly presents the considered FTTR-Business network architecture that facilitates industrial H2M collaborations, the details of our experimental studies, and the considered human's head movement prediction models. Sec.~\ref{sec4} describes the algorithm to implement the proposed HMC-DBA scheme. Sec.~\ref{sec5} evaluates different human's head movement prediction models and the HMC-DBA scheme. Finally, Sec.~\ref{sec6} summarizes the contributions of this paper.\par

\section{Review of Related Works} \label{sec2}
Enabling XR systems over state-of-the-art communication networks presents several complex challenges to the researchers from both academia and industry. Although 5G mobile networks support low-latency, high-throughput, and massive IoT services through creation of separate network slices, most XR applications demand low-latency as well as high-throughput and high-reliability \cite{xr_survey}. \textcolor{black}{Therefore, for efficient transmission of XR frames, the authors of \cite{xr_5g1} proposed the idea of \emph{frame-level integrated transmission} to formulate a resource allocation problem that maximizes the number of satisfied users subject to constraints on the datarate, reliability, and latency requirements. Nevertheless, this method does not incorporate user behavior dynamics or predict future resource demands. The authors of \cite{xr_tmp} proposed a predictive network slicing method that considers the XR frame size variations alongside the temporal characteristics of XR traffic. Although this approach enables proactive resource allocation, the traffic prediction is agnostic to user movements, making them inefficient for XR in H2M collaborations.}\par
\vspace{-0.1\baselineskip}
\textcolor{black}{To reduce the end-to-end latency and improve responsiveness of multiplayer virtual reality (VR) gaming systems, the authors of \cite{vr_mpg} proposed an optimization framework that leverages video rendering task offloading to edge servers. Nonetheless, this solution is based on simplistic traffic models assuming deterministic task sizes and uniform user movements for resource allocation. Interestingly, the authors of \cite{moyukh} proposed an AI-assisted XR provisioning scheme that virtually extends the QoE latency budget and supports more satisfied users, but assumes stable XR traffic patterns which may not generalize well in highly dynamic industrial environments. In this regard, dynamic resource allocations based on user movement-predictive techniques have shown great promise in reducing latency and bandwidth consumption. The authors of \cite{lstm_xr} applied long short-term memory (LSTM) networks to forecast users’ viewpoints for tiled streaming, enabling bandwidth savings through selective content delivery. Similarly, the authors of \cite{hmd_pred} utilized head and body motion prediction using 6 degrees-of-freedom data to pre-render scenes for mobile VR that significantly lowered motion-to-photon latency. While these approaches are effective in improving the QoE for XR in H2M collaborations, they often incur a substantial computational overhead and challenges to integrate with network resource allocation strategies.}\par
\begin{table*}[!t]
\centering
\caption{\textcolor{black}{Comparison with existing research}}
\label{table2}
\resizebox{\textwidth}{!}{%
\color{black}\begin{tabular}{c|c|l|l|cc}
\hline\hline
\textbf{Paper}             & \textbf{Application}                                                 & \multicolumn{1}{c|}{\textbf{Objective}}                                                                                                                                    & \multicolumn{1}{c|}{\textbf{Methodology}}                                                                                                                                                 & \multicolumn{2}{c}{\textbf{Performance Metric}}                                                                                                                                                            \\ \hline
\multirow{2}{*}{\cite{xr_5g1}} & \multirow{2}{*}{XR}                                                  & \multirow{2}{*}{\begin{tabular}[c]{@{}l@{}}Propose a transmission framework for \\ XR services over 5G to enhance QoE\\ of maximum number of users\end{tabular}}           & \multirow{2}{*}{\begin{tabular}[c]{@{}l@{}}Introduces a frame-level integrated \\ transmission approach where frames from \\ multiple users are optimally combined\end{tabular}}          & \multicolumn{1}{c|}{\begin{tabular}[c]{@{}c@{}}Network\\ capacity\end{tabular}}                  & \begin{tabular}[c]{@{}c@{}}$150\%$\\ increase\end{tabular}                                              \\ \cline{5-6} 
                               &                                                                      &                                                                                                                                                                            &                                                                                                                                                                                           & \multicolumn{1}{c|}{\begin{tabular}[c]{@{}c@{}}Frame\\ success rate\end{tabular}}                & \begin{tabular}[c]{@{}c@{}}$77\%$\\ increase\end{tabular}                                               \\ \hline
\multirow{2}{*}{\cite{xr_tmp}} & \multirow{2}{*}{XR}                                                  & \multirow{2}{*}{\begin{tabular}[c]{@{}l@{}}Analyze XR traffic patterns to aid \\ predictive network slicing\end{tabular}}                                                  & \multirow{2}{*}{\begin{tabular}[c]{@{}l@{}}Conducts temporal analysis of XR traffic \\ and proposes prediction models for future \\ frame sizes to optimize network slicing\end{tabular}} & \multicolumn{1}{c|}{\begin{tabular}[c]{@{}c@{}}Bandwidth\\ utilization\end{tabular}}             & \begin{tabular}[c]{@{}c@{}}$15\%$\\ increase\end{tabular}                                               \\ \cline{5-6} 
                               &                                                                      &                                                                                                                                                                            &                                                                                                                                                                                           & \multicolumn{1}{c|}{\begin{tabular}[c]{@{}c@{}}End-to-end\\ latency\end{tabular}}                & \begin{tabular}[c]{@{}c@{}}$60\%$\\ decrease\end{tabular}                                               \\ \hline
\cite{vr_mpg}                  & \begin{tabular}[c]{@{}c@{}}VR \\ Gaming\end{tabular}                 & \begin{tabular}[c]{@{}l@{}}Model and optimize wireless \\ multiplayer VR game systems \\ utilizing edge computing\end{tabular}                                             & \begin{tabular}[c]{@{}l@{}}Develops a framework that integrates edge \\ computing to manage computational tasks \\ and reduce latency in multiplayer VR gaming\end{tabular}               & \multicolumn{1}{c|}{\begin{tabular}[c]{@{}c@{}}Inter-player\\ latency\end{tabular}}              & \begin{tabular}[c]{@{}c@{}}0.2390 msec\\ (worst-case)\end{tabular}                                      \\ \hline
\cite{moyukh}                  & XR                                                                   & \begin{tabular}[c]{@{}l@{}}Improve XR service provisioning via \\ predictive AI in 5G mobile systems.\end{tabular}                                                                     & \begin{tabular}[c]{@{}l@{}}Uses AI-based prediction of XR frames to \\stretch ideal QoE latency budget while \\maintaining quality.\end{tabular}                                        & \multicolumn{1}{c|}{\begin{tabular}[c]{@{}c@{}}No. of\\ satisfied\\ users\end{tabular}}          & \begin{tabular}[c]{@{}c@{}}$180\%$\\ increase\end{tabular}                                              \\ \hline
\cite{lstm_xr}                 & VR                                                                   & \begin{tabular}[c]{@{}l@{}}Improve VR streaming quality \\ by predicting user viewpoints.\end{tabular}                                                                     & \begin{tabular}[c]{@{}l@{}}Utilizes LSTM networks to predict future \\ viewpoints, enabling efficient multi-quality \\ tiled video coding\end{tabular}                                    & \multicolumn{1}{c|}{\begin{tabular}[c]{@{}c@{}}Bandwidth\\ consumption\end{tabular}}             & \begin{tabular}[c]{@{}c@{}}$61\%$\\ decrease\end{tabular}                                               \\ \hline
\cite{hmd_pred}                & VR                                                                   & \begin{tabular}[c]{@{}l@{}}Enable low-latency mobile VR \\ by predicting user head and body \\ movements\end{tabular}                                                      & \begin{tabular}[c]{@{}l@{}}Proposes a motion prediction system that \\ anticipates user movements to pre-render and \\ stream VR content, reducing perceived latency\end{tabular}         & \multicolumn{1}{c|}{\begin{tabular}[c]{@{}c@{}}Accuracy of\\ pre-rendered\\ frames\end{tabular}} & \begin{tabular}[c]{@{}c@{}}$>99\%$\\ increase\end{tabular}                                              \\ \hline
\cite{xr_quality}              & XR                                                                   & \begin{tabular}[c]{@{}l@{}}Develop an index to evaluate the \\ transmission quality of XR services \\ in 5G networks\end{tabular}                                          & \begin{tabular}[c]{@{}l@{}}Proposes the XR Quality Index (XQI), which \\ assesses transmission quality based on factors \\ like latency, packet loss, and jitter\end{tabular}             & \multicolumn{1}{c|}{\begin{tabular}[c]{@{}c@{}}User\\ satisfaction\\ ratio\end{tabular}}         & \begin{tabular}[c]{@{}c@{}}$>68\%$ (XQI$=5$)\\ $>92\%$ (XQI$\geq4$)\\ $>98\%$ (XQI$\geq3$)\end{tabular} \\ \hline
\multirow{2}{*}{This paper}    & \multirow{2}{*}{\begin{tabular}[c]{@{}c@{}}XR\\ in H2M\end{tabular}} & \multirow{2}{*}{\begin{tabular}[c]{@{}l@{}}Design an end-to-end resource allocation\\ scheme that synchronizes human's head\\ and machine's camera movements\end{tabular}} & \multirow{2}{*}{\begin{tabular}[c]{@{}l@{}}Uses predicted human head movements to\\ synchronize camera orientations and pre-\\ allocate future bandwidth for XR traffic\end{tabular}}     & \multicolumn{1}{c|}{\begin{tabular}[c]{@{}c@{}}End-to-end\\ latency\end{tabular}}                & \begin{tabular}[c]{@{}c@{}}$>98\%$\\ decrease\end{tabular}                                              \\ \cline{5-6} 
                               &                                                                      &                                                                                                                                                                            &                                                                                                                                                                                           & \multicolumn{1}{c|}{\begin{tabular}[c]{@{}c@{}}Bandwidth\\ consumption\end{tabular}}             & \begin{tabular}[c]{@{}c@{}}$>80\%$\\ decrease (8K)\end{tabular}                                         \\ \hline
\end{tabular}%
}
\end{table*}
\setlength{\textfloatsep}{1pt}
%
\textcolor{black}{In conjunction with efficient XR traffic transmission and user movement prediction techniques, the authors of \cite{xr_quality} introduced XR Quality Index (XQI) as a comprehensive metric to evaluate the quality of XR transmission over radio access networks. This is an extremely useful metric for post-hoc diagnostics and long-term network planning for efficient XR services, but the authors did not provide with any scheme to dynamically adapt network resource allocation against applications like industrial H2M collaborations with changing network load or user movement complexity. In summary, the aforementioned works focused on either transmission efficiency, user movement prediction, or edge server-assisted XR content rendering without providing an end-to-end, predictive, and XR QoE-aware resource allocation framework. Our proposed HMC-DBA scheme, leveraging the control and management interfaces of FTTR-Business network architecture, addresses these issues to enable ideal immersive XR in H2M collaborations over a large geographic span with perfect synchronization between machine's camera and human's head movements. Table \ref{table2} presents a critical comparison between our proposed framework and the aforementioned works.}

\section{System Model} \label{sec3}
\subsection{XR in H2M Collaborations over FTTR-Business Network} \label{sec3.1}
This work considers deployment scenario of H2M collaborations over FTTR-Business network in an industrial scenario where multiple humans from the office building are performing some tasks at a remote manufacturing site building by interacting with some machines. As shown in Fig. \ref{hcs}, the main OLT hosts an edge-AI server where the H2M application and AI-based prediction task are done to support the OLT for dynamic resource allocation for the 50G-EPON (maximum span of 20 km). The ONUs are usually connected with a WAP for providing multiple-access to multiple users. Some of the ONUs can be referred to as MFUs which serves like an OLT for the SFUs in the FTTR segment. These SFUs are connected by a 10G-EPON or 50G-EPON (maximum span of 20 m) to bring services to the users in different rooms via WAPs like WiFi 6/6E (9.6 Gbps), WiFi 7 (46 Gbps), or 5G (10 Gbps).\par
\textcolor{black}{Note that the two-stage FTTR architecture still follows the ONU management and control interface protocol of ITU-T G.988 for the TDM-PON \cite{ITU-T_G.988}. Practical challenges like time synchronization between the OLTs and ONUs are achieved by ranging mechanism, i.e., measuring the round-trip time between the nodes at the initialization. Moreover, the effects of clock drifts and latency in control messages are mitigated by guard interval between received data from different ONUs. In addition, the FTTR networks provides the provision for designing a centralized fiber-WiFi integration framework facilitating data transmission in the uplink and downlink without typical issues like channel contention, packet loss, interference, signal fading, or bandwidth variability that might affect the QoE of XR in industrial H2M collaborations \cite{fttr_jocn}. The newly introduced control and management channels (separate from data channels) between OLT to MFUs and between MFUs and SFUs can be used to continuously monitor network status and accordingly provide control commands to update resource allocation strategies. Each MFU can activate/deactivate the management channel from the main OLT to each of the SFUs connected with it. The MFUs do not process the management and control messages for the SFUs from the main OLT but aggregate and transparently transmit.} 

\subsection{Experimental Study on XR and HMD Traffic} \label{sec3.2}
For our experimental setup, we created a WiFi interface between a lab computer and a commercially available educational purpose robot, \emph{JD Humanoid} from EZ-Robot \cite{jd_robot}. \textcolor{black}{An application software \emph{ARC by Synthiam} is used to control and orchestrate the robot to perform various tasks typically performed during industrial H2M collaborations like picking up some object, placing an object on a rack, and tightening a screw with a screwdriver \cite{ind_act}. This application also coordinates the robot's camera and the human's head movements in synchronization such that they both see the same scene.}\par
\begin{figure}[!t]
    \centering
    \subfloat[Gaussian(33.13, 3.08)]{%
    \includegraphics[width=0.5\columnwidth]{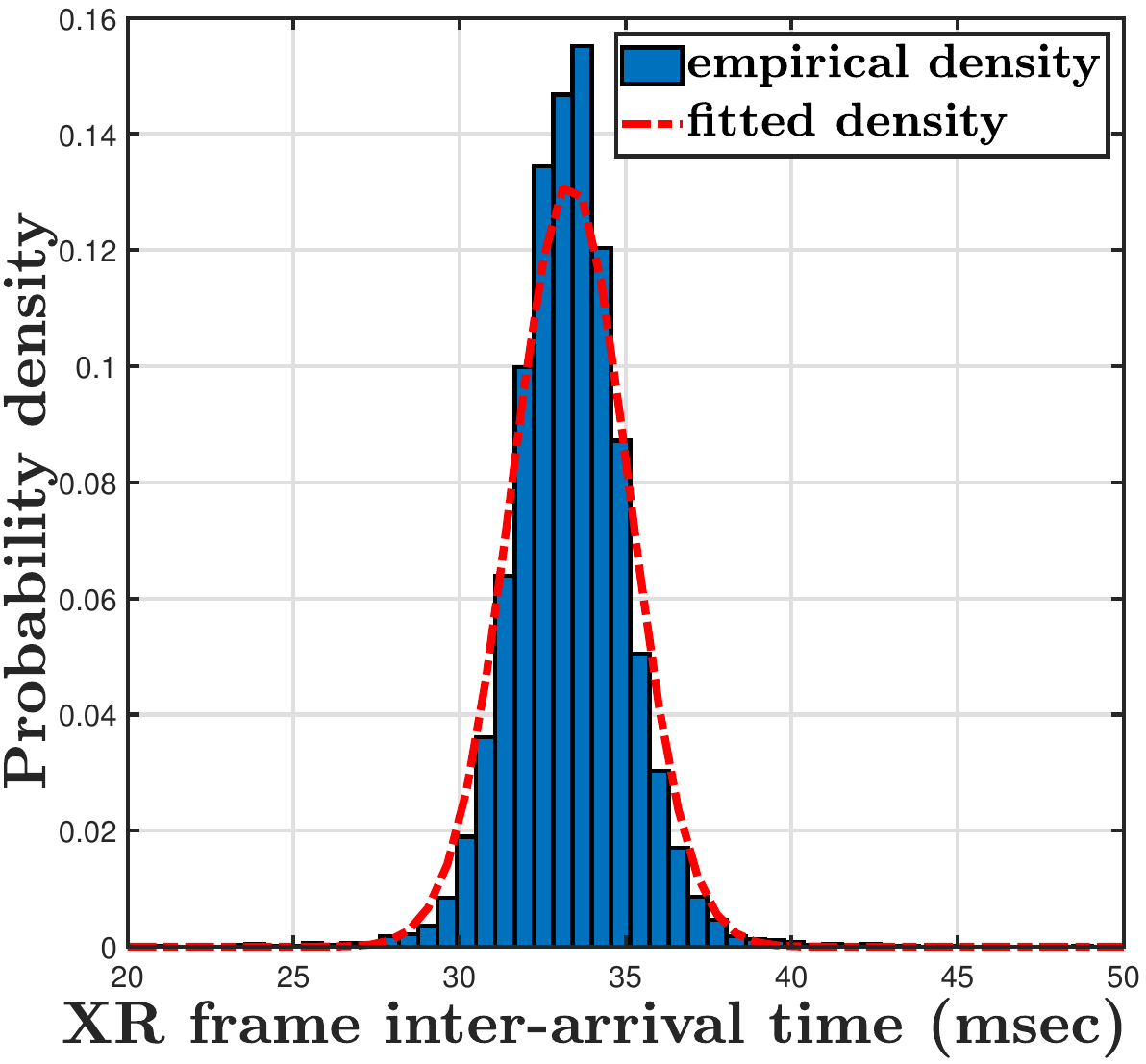}\label{exp1_1}%
    }
    \subfloat[Gamma(0.8839, 33.6439)]{%
    \includegraphics[width=0.5\columnwidth]{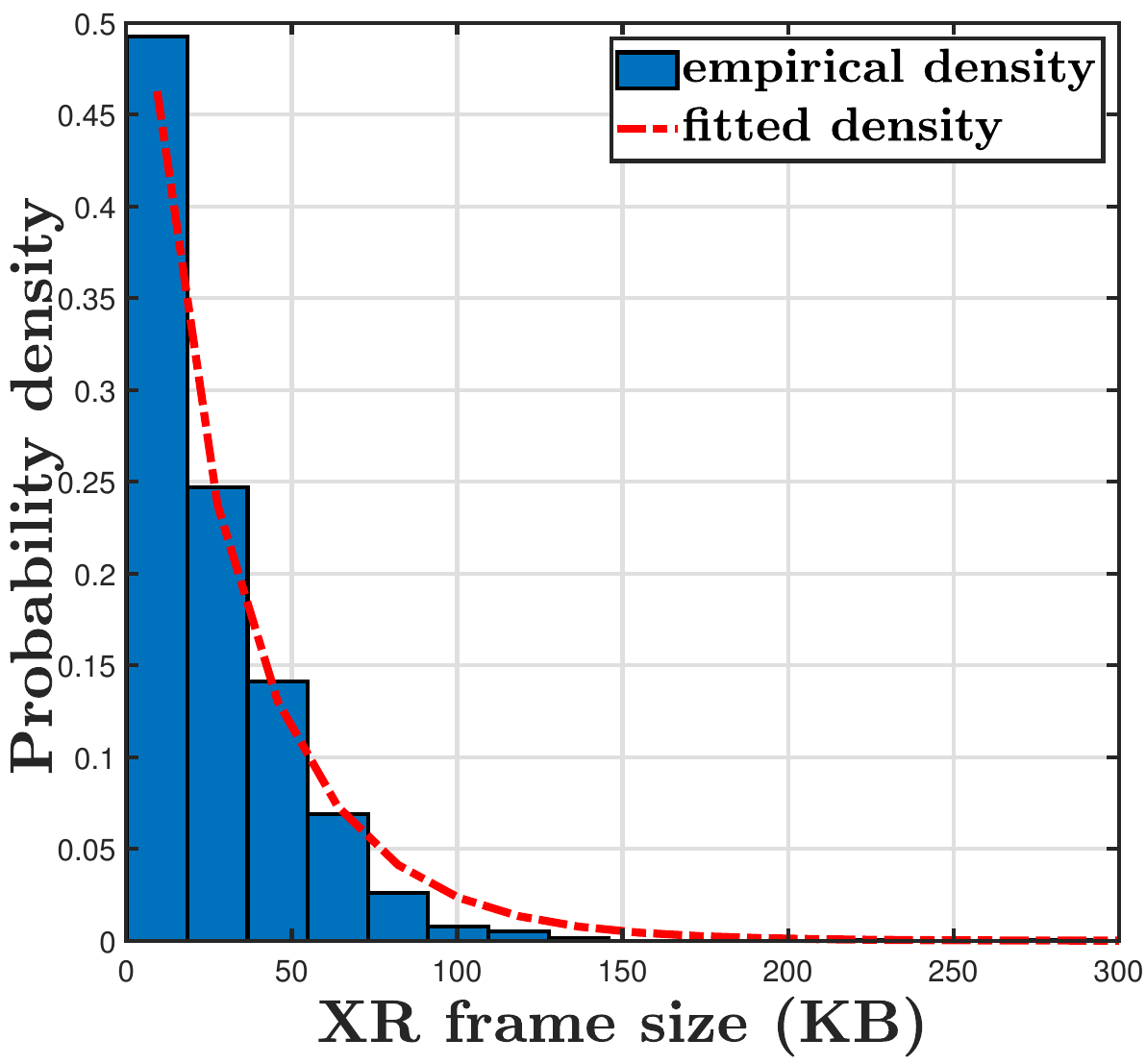}\label{exp1_2}%
    }

    \caption{\textcolor{black}{Experimental XR traffic characteristics: (a) frame inter-arrival times fitted with a Gaussian(33.13, 3.08) distribution (Mean = 33.13 msec, STD = 1.76 msec), (b) frame sizes fitted with a Gamma(0.8839, 33.6439) distribution (Mean = 29.74 KB, STD = 100 KB).}}
    \label{xr_data}
\end{figure}
\setlength{\textfloatsep}{5pt}
%
\textcolor{black}{\textbf{(a) XR traffic model:} The best resolution of this camera is $640\times 480$ (12 pixels/degree) with frame rate = 30 fps and raw datarate $\approx 18.432$ Mbps. Although the content-resolution is much lower than different XR QoE requirements mentioned in Table \ref{table1}, the primary benefit of using this setup is that we can study the XR traffic characteristics by performing the same tasks during an industrial H2M collaboration in our lab environment. As the place of actions of XR in H2M collaborations are mostly inside some building premises \cite{itu_fttr}, the wireless channel conditions of the lab experiments are very similar to the real-world applications. From the recorded timestamps of the XR traffic, we observed that it is \emph{pseudo-periodic} in nature, i.e., it appears to be repeating itself at regular intervals like a periodic function, but does not follow an exact deterministic pattern. If $x(t)$ is a periodic signal with period $T$, then, $x(t) = x(t + T), \forall t$, but a pseudo-periodic signal $y(t)$ with average period $T$ can be defined as $y(t)\approx y(t+T+\epsilon(t)), \forall t$, where $\epsilon(t)$ is a slowly varying function or jitter following any arbitrary distribution.}\par
\textcolor{black}{Recently, the 3GPP XR study group \cite{3gpp_xr} has provided with a standardized model for XR traffic which mentions that the jitter is expected to follow a truncated Gaussian distribution with Mean = 0 msec, standard deviation (STD) = 2 msec, and [Min, Max] $\leq$ [-5, 5] msec. Fig.~\ref{exp1_1} shows that our observed XR traffic fits well with a Gaussian distribution with Mean = 33.13 msec, STD = 1.76 msec, hence consistent with the standard requirements. The frame size is expected to follow a Truncated Gaussian distribution with Mean = (average datarate)/(frame rate $\times$ 8) bytes and [STD, Min, Max] = $[10.5\%, 50 \%, 150\%]$ of Mean. For example, if we consider 2K with 60 fps (refer to Table \ref{table1}), then the average frame size will be (40 Mbps/60 fps) $\approx$ 83.33 KB. Nevertheless, this is applicable to the raw data, but we observed from our experiments that the effective frame size followed a Gamma(0.8839, 33.6439) distribution with Mean = $29.74$ KB and STD = 100 KB, as shown in Fig.~\ref{exp1_2}. After a thorough investigation, we realized that the primary reason behind this anomaly is the dependence of effective datarate on frame compression techniques using inter-frame correlations. \emph{While performing tasks in collaboration with a robot, the humans do not move their heads continuously, but intermittently like a random on-off process}. This makes the inter-frame correlations very high, which in turn, greatly contributes to reduce XR frame size and effective data rate. Note that usually each XR frame are segmented into multiple Ethernet packets with payload $\leq 1500$ bytes and transmitted as a burst over networks.}\par
\begin{figure}[!t]
    \centering
    \subfloat[Gamma(51.844, 0.273)]{%
    \includegraphics[width=0.5\columnwidth]{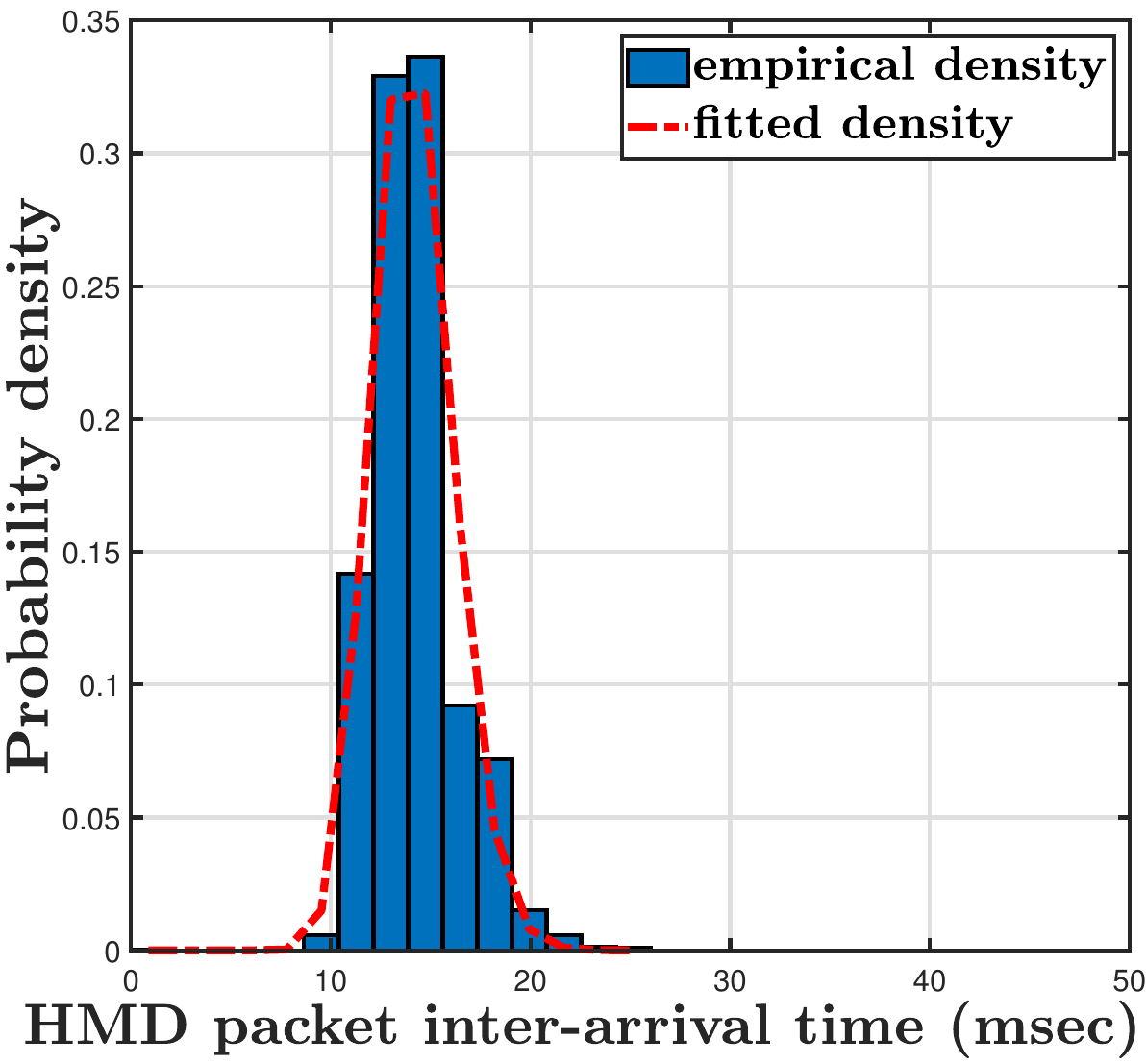}\label{exp2_1}%
    }
    \subfloat[An instance of HMD data]{%
    \includegraphics[width=0.5\columnwidth]{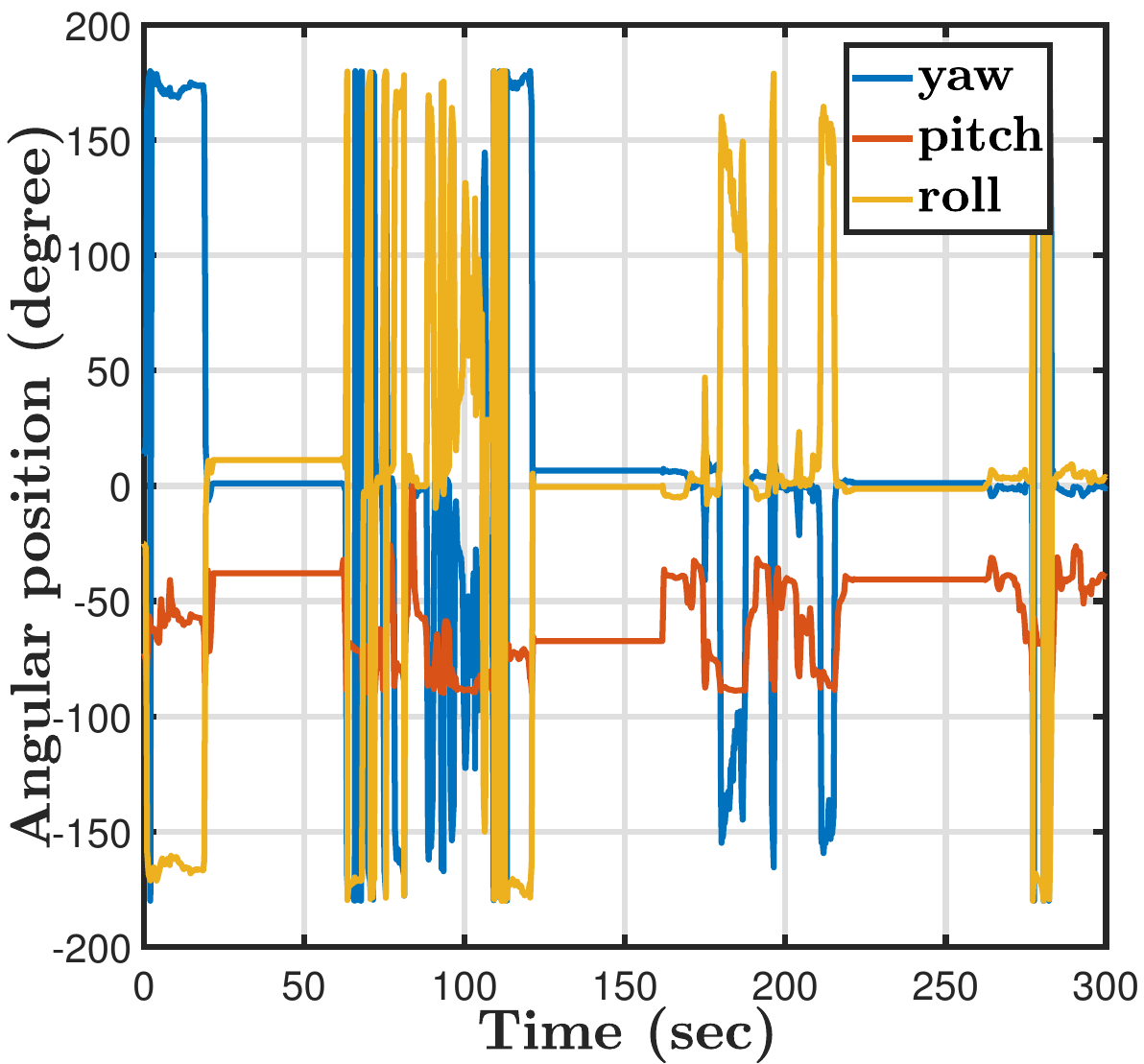}\label{exp2_2}%
    }

    \subfloat[Autocorrelation and partial autocorrelation of yaw, pitch, and roll]{%
    \includegraphics[width=\columnwidth]{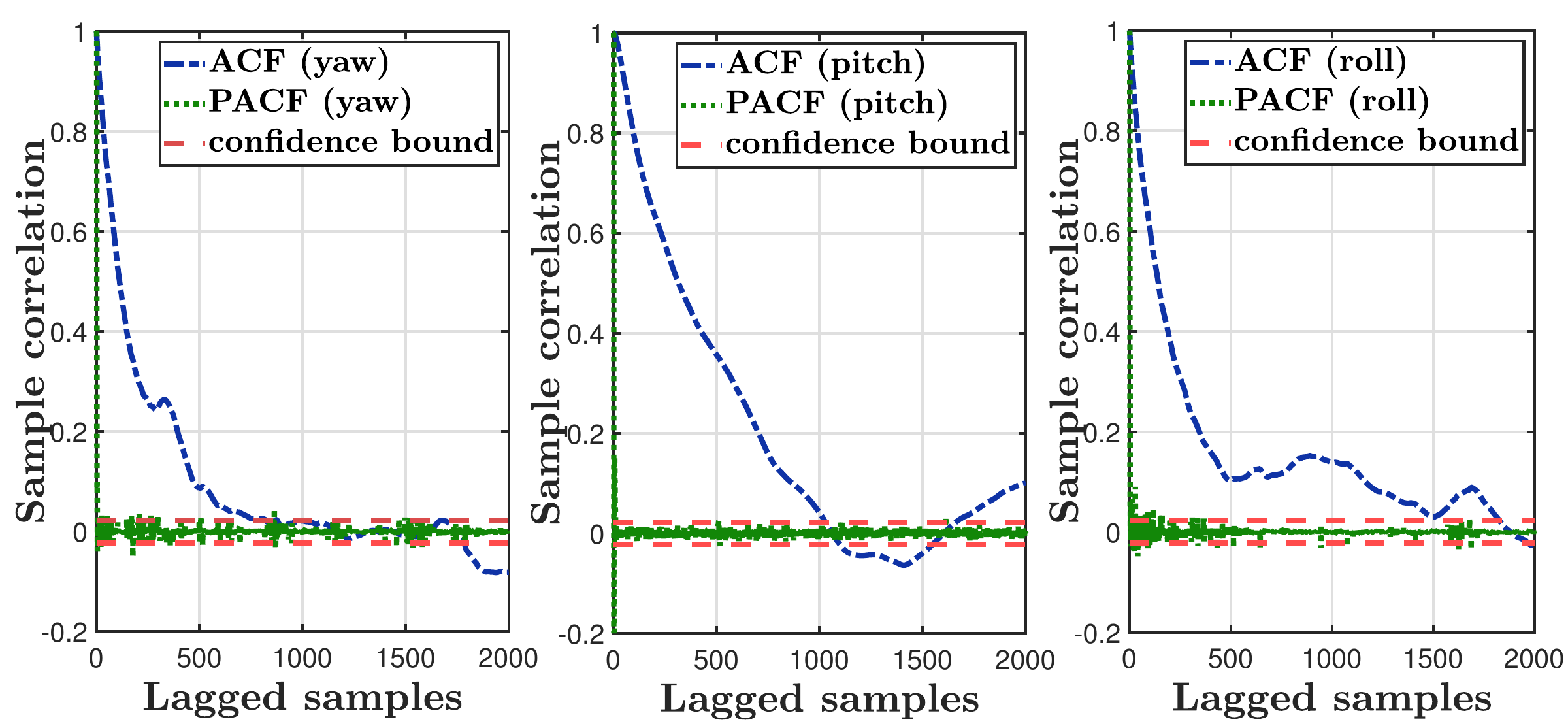}\label{exp3}%
    }

    \caption{\textcolor{black}{Experimental HMD data characteristics: (a) empirical and fitted [Gamma(51.844, 0.273)] distribution of packet inter-arrival times, (b) sequence of yaw, pitch, and roll values, and (c) ACF and PACF plots of yaw, pitch, and roll (100 samples $\approx 1$ sec).}}
    \label{hmd_data}
\end{figure}
\setlength{\textfloatsep}{5pt}
%
\textbf{(b) HMD traffic model:} We observed that the inter-arrival times of packets containing HMD orientations are also pseudo-periodic following a Gamma(51.844, 0.273) distribution with Mean = $14.13$ msec and STD = 1.96 msec, as shown in Fig. \ref{exp2_1}. In general, during any practical H2M collaboration, the human's head movement data do not show any trend or seasonality (varies from operation to operation), but only random cycles of swing between highest and lowest values, as shown in Fig. \ref{exp2_2}. Theoretically, this type of time-series data are referred to as \emph{stationary} because these do not have any predictable patterns in the long run. In Fig. \ref{exp3}, we plot the sample autocorrelation functions (ACFs) and partial autocorrelation functions (PACFs) for yaw, pitch, and roll components. These ACFs show a very slow degradation of sample correlation values and approach the 95\% confidence bound around a lag of 1000 samples, but do not completely settle down within the confidence bounds. However, the PACF plots rapidly settle down within the 95\% confidence bound. This implies that designing an efficient prediction model for these sequences is not straight-forward as it has long-term memory. Hence, we need to compare well-known sequence prediction models such as persistence, moving average, regression, ARIMA, and bidirectional LSTM (BiLSTM) networks to find the best predictor of human's head movements.\par
\begin{figure*}[!t]
\centering
\includegraphics[width=0.97\textwidth]{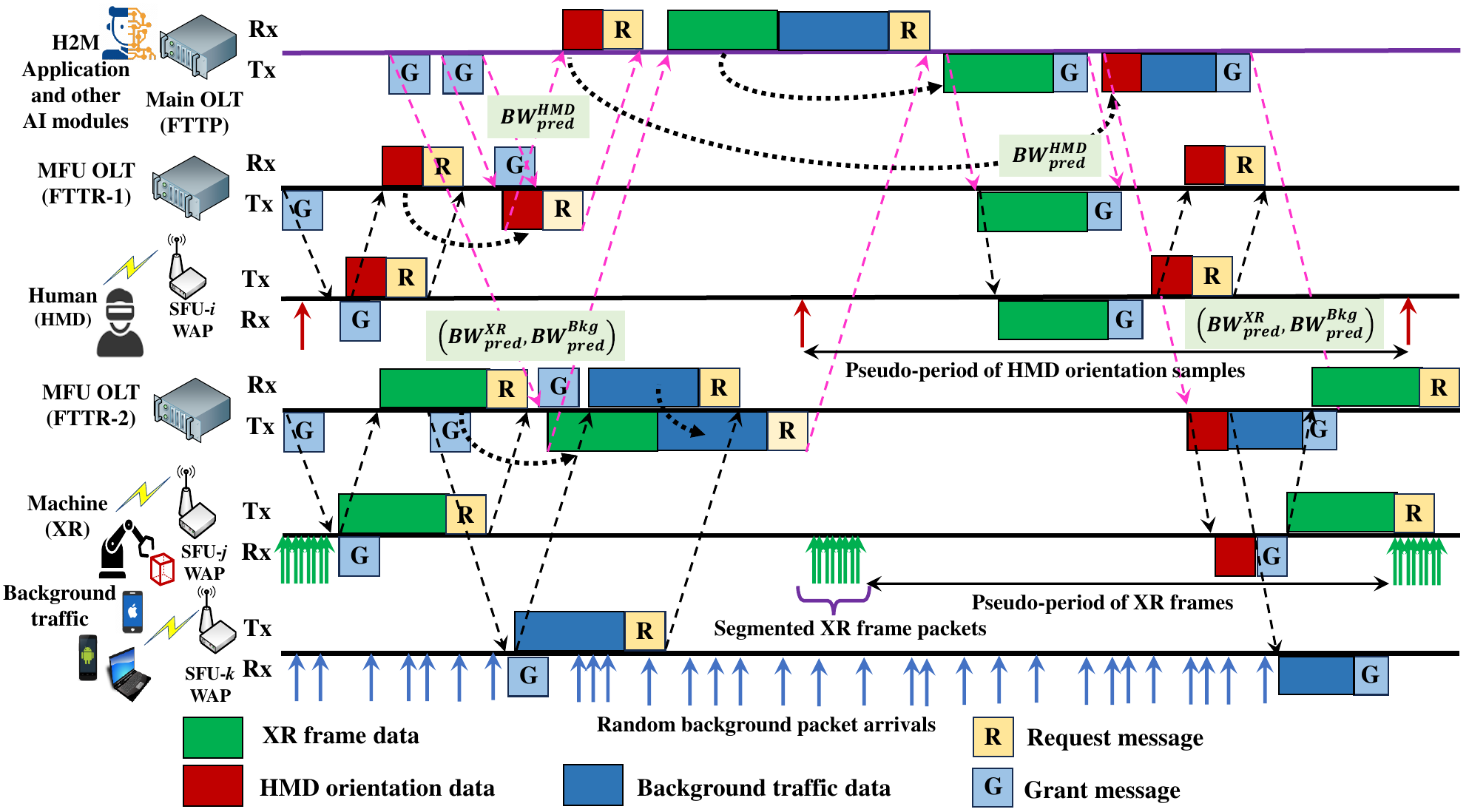}
\caption{An illustration of HMC-DBA scheme with ONU-WAPs connected to human (HMD), machine (XR), and background traffic.}
\label{fttr}
\end{figure*}
\setlength{\textfloatsep}{1pt}
\textbf{(c) Estimation of XR frame size from HMD orientations:} The HMD orientations can be recorded either in Euler angles viz., yaw, pitch, roll, or normalized quaternions \cite{quat2euler}. We denote yaw by $\theta_y \in [-180\degree,180\degree]$, pitch by $\theta_p \in [-90\degree,90\degree]$, and roll by $\theta_r \in [-180\degree,180\degree]$. They can be converted to normalized quaternion coordinates $(q_w,q_x,q_y,q_z)$ such that $\lVert \bm{q} \rVert = \sqrt{q_w^2 + q_x^2 + q_y^2 + q_z^2} = 1$ as shown in \cite{eul_quat}. Moreover, the angular distance between two quaternions $\bm{q}_1 = (q_{w1},q_{x1},q_{y1},q_{z1})$ and $\bm{q}_2 = (q_{w2},q_{x2},q_{y2},q_{z2})$ can be evaluated as $\theta_d = 2\arccos(\text{Real}\{\bm{q}_1 \otimes \bm{q}_2^*\})$. The \emph{average speed of head movement} $(\phi)$ in degree/sec with $n$ samples is calculated as follows:
\begin{align}
    \phi = \frac{1}{n-1}\sum_{i=1}^{n-1} \frac{\theta_d(\bm{q}_i,\bm{q}_{i-1})}{\Delta \tau_i}, \label{eq01}
\end{align}
where, $\theta_d(\bm{q}_i,\bm{q}_{i-1})$ indicates the angular shift between consecutive samples and $\Delta \tau_i$ indicates inter-sample times.\par
\textcolor{black}{Interestingly, XR frame size can also be estimated using human head movements. Let $F = W \times H$ denote the frame dimension (pixels) where $W$ denotes horizontal resolution and $H$ denotes vertical resolution. Moreover, $f$ (fps) denotes frame rate, $d$ (bytes/pixel) denotes color depth, and $\gamma$ denotes FOV (degree). Therefore, the raw datarate $R_{raw}$ can be calculated as $R_{raw} = F \times d \times f$ (bytes/sec). Now, the pixel shift in consecutive XR frames based on the angular shift $\theta_d(t)$ at time instant $t$ is calculated by $\delta x(t) = \frac{F\times\theta_d(t)}{\gamma}$. As angular shift and pixel shift are directly proportional, we can calculate the effective frame size by $F_{eff}(t) = R_{eff}(t)/f$ and $R_{eff}(t)$ is derived as follows:
\begin{align}
    R_{eff}(t) = F \times d \times f \times \left(1-\exp\left\{-k\frac{F\times\theta_d(t)}{\gamma}\right\}\right), \label{eq02}
\end{align}
where, $k$ is a sensitivity constant depending on the inter-frame correlation-based compression technique and the decaying exponential function imposes saturation in datarate increment due to an angular shift beyond FOV. This model also justifies our observations on XR frame size distribution.}\par

\subsection{HMD Movement Prediction} \label{sec3.3}
In \emph{persistence method}, the last recorded HMD orientation is used as the immediately predicted value, assuming a static model. This model works reasonably well for slowly varying sequences but prediction error increases with higher prediction horizons as. In \emph{moving average}, a window of past samples is chosen to take their average and considered as the predicted value \cite{mvg_avg}. Performance of this model is dependent on the choice of moving window because a larger window yields poor performance against fast varying sequences. The \emph{ARIMA model} works best against data sequences that can be made stationary by differencing to eliminate the non-stationarity of the mean function and its short-term random time patterns remains the same statistically \cite{arima}. \textcolor{black}{The \emph{LSTM} and \emph{BiLSTM models} are special type of deep neural networks capable of predicting long-term non-linear time-series data like human head movements and several times more efficient to predict sudden transitions than persistence method \cite{lstm_ref}. If a window of past samples are provided as feature inputs and the network is trained over a significantly large dataset, then prediction accuracy even for long horizons can be expected to be satisfactory. In particular, BiLSTM networks are more efficient than standard LSTM networks as they process the input sequence in both forward and backward directions. This makes the network capable of accessing both past and future contexts during training and capture subtle temporal dependencies, leading to an improved prediction accuracy. For training purpose, we configured the BiLSTM network with hyperparameters such as the number of layers = 3, hidden units per layer = 200, dropout rate =  0.3 (to prevent overfitting), and optimized using stochastic gradient descent with momentum technique. Moreover, we minimized the mean square error loss function over 50 epochs with early stopping to prevent over-training.}
\section{Human-Machine Coordinated DBA} \label{sec4}
In the considered FTTR-Business architecture, there are FTTR segments (10G or 50G-EPONs) cascaded after a FTTP segment (50G-EPON). The SFUs in the FTTR segment are usually connected to WiFi 6/6E, WiFi 7 or 5G WAPs. As both the humans and machines are operating in in-premise environments, usually the channel conditions are very good with very high SINR between the WAPs and devices. Moreover, due to centrally coordinated transmission from all the WAPs, packet collisions and channel contentions can be avoided very efficiently \cite{fttr_jocn}. In Fig.~\ref{fttr}, we illustrate the working principles of the HMC-DBA scheme with timeline. We show that SFU-$i$ corresponding to FTTR-1 is supporting a human collaborator and SFU-$j$ corresponding to FTTR-2 is supporting a machine collaborator. There are more number of such SFUs, but for the sake of illustration, we are showing only a pair of human and machine collaborators. \textcolor{black}{Besides this, some SFUs are supporting miscellaneous devices contributing to random background traffic. The aggregated background traffic from each SFU can be upto several Gbps with packet inter-arrival times following exponential or Pareto distributions. The functions of different nodes are step-by-step summarized below.}\par
\textcolor{black}{\textbf{(a) SFUs requesting bandwidth:} As HMD orientations from humans, XR frames from machines, and background traffic from other devices arrive at the respective SFUs over wireless interfaces, they send a request message $R = BW_{req}$ piggybacked with their uplink (UL) data to the connected MFU in their respective FTTR for granting uplink bandwidth in the next polling cycle.}\par
\textcolor{black}{\textbf{(b) MFUs granting bandwidth:} If the baseline limited service DBA (LS-DBA) scheme was employed, then the MFUs would have send a grant message $G$ piggybacked to the downlink (DL) data with bandwidth $\min\{BW_{req}^{HMD}, BW_{max}\}$, $\min\{BW_{req}^{XR}, BW_{max}\}$, and $\min\{BW_{req}^{Bkg}, BW_{max}\}$, respectively to humans, machines, and background devices. Note that $BW_{max}$ indicates the maximum amount of bandwidth can be allocated to each SFU in one polling cycle. However, with our proposed HMC-DBA scheme, the MFUs will send grants $\min\{BW_{pred}^{HMD}, BW_{max}\}$, $\min\{BW_{pred}^{XR}, BW_{max}\}$, and $\min\{BW_{pred}^{Bkg}, BW_{max}\}$ respectively. Therefore, instead of granting the SFUs with only the requested bandwidth, some excess bandwidth is granted based on their traffic arrival predictions \cite{lihua_iotj}. This reduces the queueing time of traffic that arrives after the SFUs send their request messages.}\par
\textcolor{black}{\textbf{(c) SFUs transmitting data:} After receiving the grant messages, immediately the SFUs transmit their queued data to the MFU. The SFUs also transmit information about the WAPs and connected devices to the MFUs and main OLT over the control interfaces that are used to generate contention-free bandwidth allocation maps for all devices.}\par
\textcolor{black}{\textbf{(d) MFUs transmitting data:} All the MFUs in the FTTP segment send request messages to the main OLT after receiving from their respective SFUs. Following the same scheme as above, the main OLT sends the grant messages to the MFUs so that they can broadcast the DL data to the SFUs and transmit the aggregated UL traffic to the main OLT.}\par
\begin{figure}[!t]
\centering
\includegraphics[width=\columnwidth,keepaspectratio]{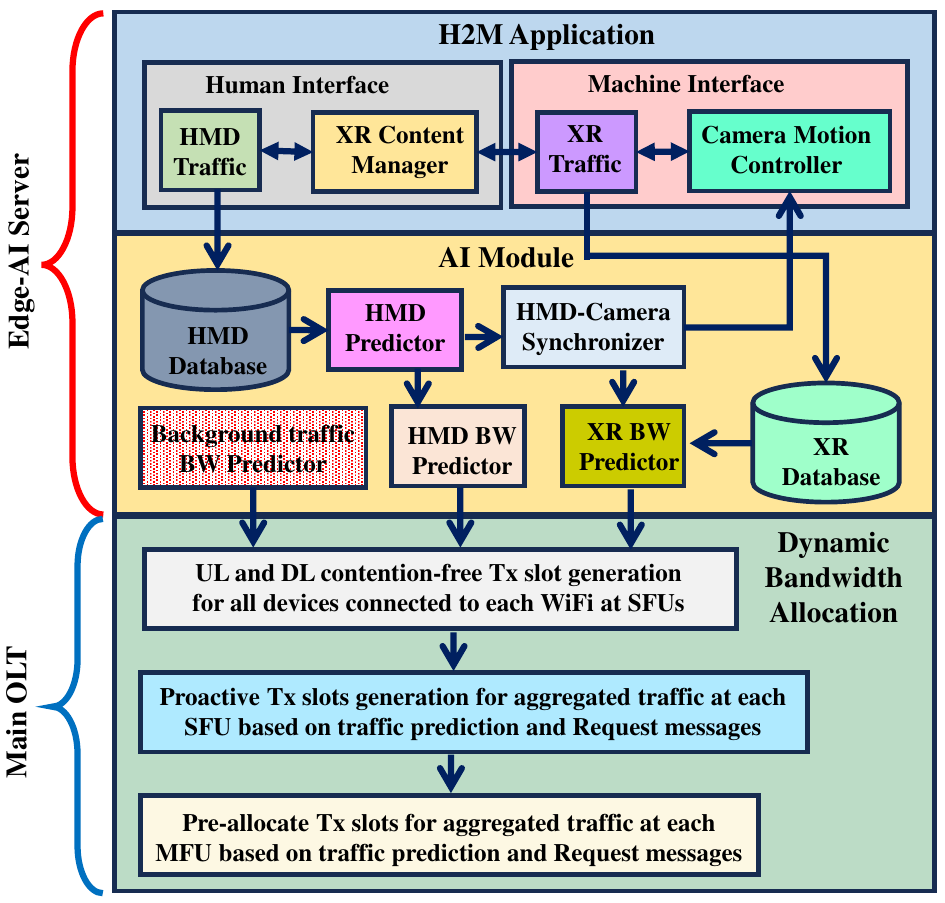}
\caption{\textcolor{black}{A schematic of the HMC-DBA scheme implemented at the Edge-AI server and main OLT of FTTR-Business networks.}}
\label{arch}
\end{figure}
\setlength{\textfloatsep}{5pt}
\textcolor{black}{\textbf{(e) Traffic prediction at the Edge-AI server:} Fig.~\ref{arch} shows the implementation of our proposed HMC-DBA scheme. After receiving the UL traffic, the main OLT forwards the HMD traffic to the human interface and the XR traffic to the machine interface of the H2M application running at the Edge-AI server. The HMD traffic are passed on to the \emph{AI module} for storing them in the HMD database and predict future human head movements as discussed in Sec.~\ref{sec3.3}. The HMD-Camera synchronizer uses this predictions further to pre-orient machine's camera through the camera motion controller according to the gap between rotational speeds of human head and machine's camera. These human head rotations are also used to predict future HMD bandwidth requirements. The XR content manager at the human interface checks if the QoE requirements of the XR frames are satisfied and augment more contents (if required). The XR traffic is also passed on to the AI module for storing them in the XR database. The XR bandwidth predictor uses the bandwidth demands from recent past and predicted human head rotations to estimate future XR bandwidth requirements. The AI module also predicts the future bandwidth requirements for the background traffic.}\par
\textcolor{black}{\textbf{(f) Bandwidth allocation by the main OLT:} Gathering all the bandwidth predictions, the main OLT first generates transmission (Tx) slots for all the UL and DL traffic to the humans, machines, and background devices connected to the SFUs. The grant messages to the SFUs from their respective MFUs will be scheduled in accordance to the UL slots of the WAPs such that the queueing time is minimum. Note that a guard time (usually 1-2 $\mu$sec) will be observed to overcome synchronization and clock drift issues. Similarly, the main OLT will send grant messages to the MFUs such that the aggregated traffic from all the SFUs experience a minimum queueing time. Although there is no standard defined maximum polling cycle duration $T_{poll}$ of a TDM-PON, in practice, it is chosen based on the QoE requirements of the traffic and the rule of thumb,
\begin{align}
    T_{poll} \leq 2\times N_{onu} \times T_{RTT} + T_{g} \times (N_{onu}-1), \label{eq03}
\end{align}
where, $N_{onu}$ denotes the number of ONUs, $T_{RTT}$ denotes the maximum round-trip-time between OLT and ONUs, and $T_{g}$ denotes the guard time. The steps to implement HMC-DBA scheme are summarized in Algorithm \ref{alg1}.}\par
\begin{algorithm}[t!]
\caption{Algorithm for HMC-DBA} \label{alg1}
\hspace*{\algorithmicindent} \textbf{Parameters:} \\
\hspace*{\algorithmicindent} $N_{sfu}$: Maximum number of SFUs in each FTTR \\
\hspace*{\algorithmicindent} $T_{poll}^{[i]}$: Starting time of $i$-th polling cycle \\
\hspace*{\algorithmicindent} $B_k^{[i]}$: Pending packet size in the buffer of ONU $k$\\
\hspace*{\algorithmicindent} $G_k^{[i]}$, $R_k^{[i]}$: Grant and Request message for ONU $k$\\
\hspace*{\algorithmicindent} $T_{XR}$: Average period of XR frame generation \\
\hspace*{\algorithmicindent} $T_{HMD}$: Average period of HMD frame generation \\
\hspace*{\algorithmicindent} $R_{FTTR}$: Datarate of the FTTR physical link (Bytes/s)\\
\hspace*{\algorithmicindent} $T_{FTTR}^{poll}$: Duration of maximum polling cycle for FTTR\\
\hspace*{\algorithmicindent} $RTT_k$: Round-trip time between OLT and ONU $k$\\
\hspace*{\algorithmicindent} $N_{max}$: Maximum number of timeslots
\begin{algorithmic}[1]
\While{current timeslot $i \leq N_{max}$ }
    \State \textbf{\texttt{// At the SFUs in each FTTR}}
    \For{SFU $k \leftarrow 1$ \textbf{to} $N_{sfu}$}
        \If{grant message $G_k^{[i]} > 0$ is received}
            \State Transmit packets from buffer upto $G_k^{[i]}$;
            \State Piggyback the request message $R_k^{[i]} = B_k^{[i]}$;
        \EndIf
    \EndFor
    \State \textbf{\texttt{// At the MFUs (as OLT of FTTRs)}}
    \State Save the values of all received requests $R_k^{[i]}$;
    \State Calculate $BW_{max} = \frac{(T_{FTTR}^{poll}-N_{sfu}\times T_g)}{N_{sfu}} \times R_{FTTR}$;
    \If{$T_{poll} \approx T_{XR}$ \textbf{and} SFU $k$ serves XR}
        \State Generate $G_k^{[i+1]} = \max\{BW_{pred}^{XR}, BW_{max}\}$;
    \ElsIf {$T_{poll} \approx T_{HMD}$ \textbf{and} SFU $k$ serves HMD}
        \State Generate $G_k^{[i+1]} = \max\{BW_{pred}^{HMD}, BW_{max}\}$;
    \ElsIf {SFU $k$ serves Background traffic}
        \State Generate $G_k^{[i+1]} = \max\{BW_{pred}^{Bkg}, BW_{max}\}$;
    \EndIf
    \State Generate grant messages with $G_k^{[i+1]}$ for all SFU $k$;
    \State The next grant transmission times are given by
    \State $T_k^{[i+1]} = T_{poll}^{[i]}+\frac{G_k^{[i+1]}}{R_{FTTR}}+T_g+RTT_k - RTT_{(k-1)};$
    \State Forward downlink data to the SFUs with $G_k^{[i+1]}$;
    \State \textbf{\texttt{// At the MFUs (as ONUs of FTTP)}}
    \State Forward the aggregated uplink data from SFUs to
    \Statex \hspace*{\algorithmicindent}the FTTP main OLT;
    \State \textbf{\texttt{// At FTTP OLT with H2M application}}
    \If{HMD orientation received}
        \State Observe the gap between human's head and 
        \Statex \hspace*{\algorithmicindent}\hspace*{\algorithmicindent}machine's camera rotation speed;
        \State Decide the required prediction horizon;
        \State Predict the next HMD orientation using BiLSTM;
        \State Generate camera re-orientation command to be
        \Statex \hspace*{\algorithmicindent}\hspace*{\algorithmicindent}transmitted to the corresponding machine;
        \State Estimate $BW_{pred}^{XR}$ following Sec.~\ref{sec3};
    \ElsIf{XR frames received}
        \State Augment the XR frame contents if required;
        \State Save XR frame size for future $BW_{pred}^{XR}$ predictions;
    \ElsIf{Background data is received}
        \State Estimate $BW_{pred}^{Bkg}$ using known AI methods \cite{ml_ewon};
    \EndIf
    \State Generate data transmission slots for all nodes;
\EndWhile
\end{algorithmic}
\end{algorithm}
\setlength{\textfloatsep}{5pt}
\textcolor{black}{\textbf{(g) Prediction error handling:} Note that usually the XR frame size is much bigger than standard Ethernet packet size. Hence they are segmented into multiple packets and transmitted together as a burst \cite{xr_tmp}. Considering the jitter of XR and HMD traffic, the main OLT and MFUs will start to provide grants for XR and HMD devices in polling cycles slightly ahead, say $(\text{Mean} - \text{STD})$ of traffic arrival at each period and will stop after the data packets are received with request message $R = 0$. If $R > 0$, then further grants are given in the next polling cycles to compensate the effect of under-prediction. Let us denote the predicted frame size by $F_{pred}$ and the actual frame size by $F_{act}$ such that $F_{pred} = F_{act} + \rho_{XR}$, where $\rho_{XR}$ is the prediction error. Additionally, for the packets in each XR frame, we denote the waiting latency in SFU until grant message by $T_{w}^{S}$, the queueing latency until next polling cycles due to insufficient grant size by $T_{Q}^{S}$, and the transmission time in the FTTR segment by $T_{tx}^{S}$. Therefore, the total UL latency of these packets are given by,
\begin{align}
    \mathcal{L}_{XR} = T_{w}^{S} + T_{Q}^{S}(\rho_{XR}) + T_{tx}^{S}. \label{eq04}
\end{align}
\hspace*{1em}At over-prediction condition, $T_{w}^{S} \approx 0$ and $\rho_{XR} \geq 0$, $T_{Q}^{S}(\rho_{XR}) = 0$, but at under-prediction condition, $T_{w}^{S} = T_{poll}$, $\rho_{XR} < 0$ and
\begin{align}
     T_Q^{S}(\rho_{XR}) = \left\lceil{\frac{|\rho_{XR}|}{BW_{max}}}\right\rceil T_{poll}. \label{eq05}
\end{align}
\hspace*{1em}Substituting this in (\ref{eq04}), we get the latency as,
\begin{align}
    \mathcal{L}_{XR} = T_{poll} + \mathds{1}_{\{\rho_{XR} < 0\}} \times \left\lceil{\frac{|\rho_{XR}|}{BW_{max}}}\right\rceil T_{poll} + T_{tx}^{S}. \label{eq06}
\end{align}
\hspace*{1em}Next, we can derive the average packet latency $\mathbb{E}[\mathcal{L}_{XR}]$ with respect to the probability distributions of XR frame prediction error $\rho_{XR}$ and packet size. Considering jitter as the variance of latency, we can derive it as follows:
\begin{align}
    \mathcal{J}_{XR} = \text{Var}\left\{T_{poll} \left(1+\sum_{p=1}^{\infty} {\mathds{1}_{\{\rho_{XR}<-p\cdot BW_{max}\}}}\right)\right\}, \label{eq07}
\end{align}
where, $p$ denotes the number of polling cycles required to transmit $\rho_{XR}$. These equations can be further used to calculate the average latency and jitter of XR over the FTTP segment. The same process is also applicable to the packets containing HMD and background traffic.}

\begin{figure*}[!t]
  \centering
  \subfloat[Prediction of yaw]{%
    \includegraphics[width=0.333\textwidth]{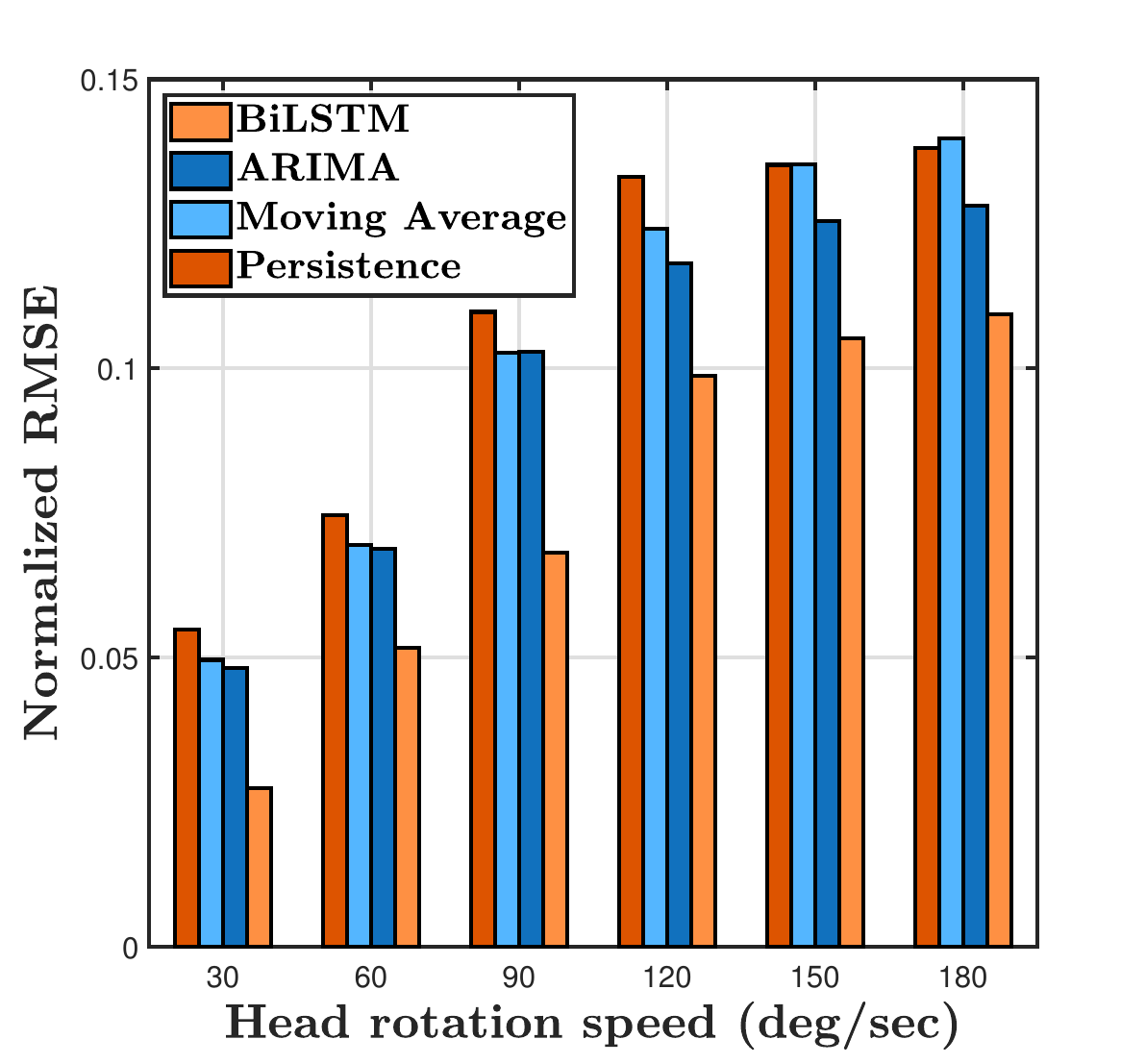}\label{rmmse_yaw6}%
  }
  \subfloat[Prediction of pitch]{%
    \includegraphics[width=0.333\textwidth]{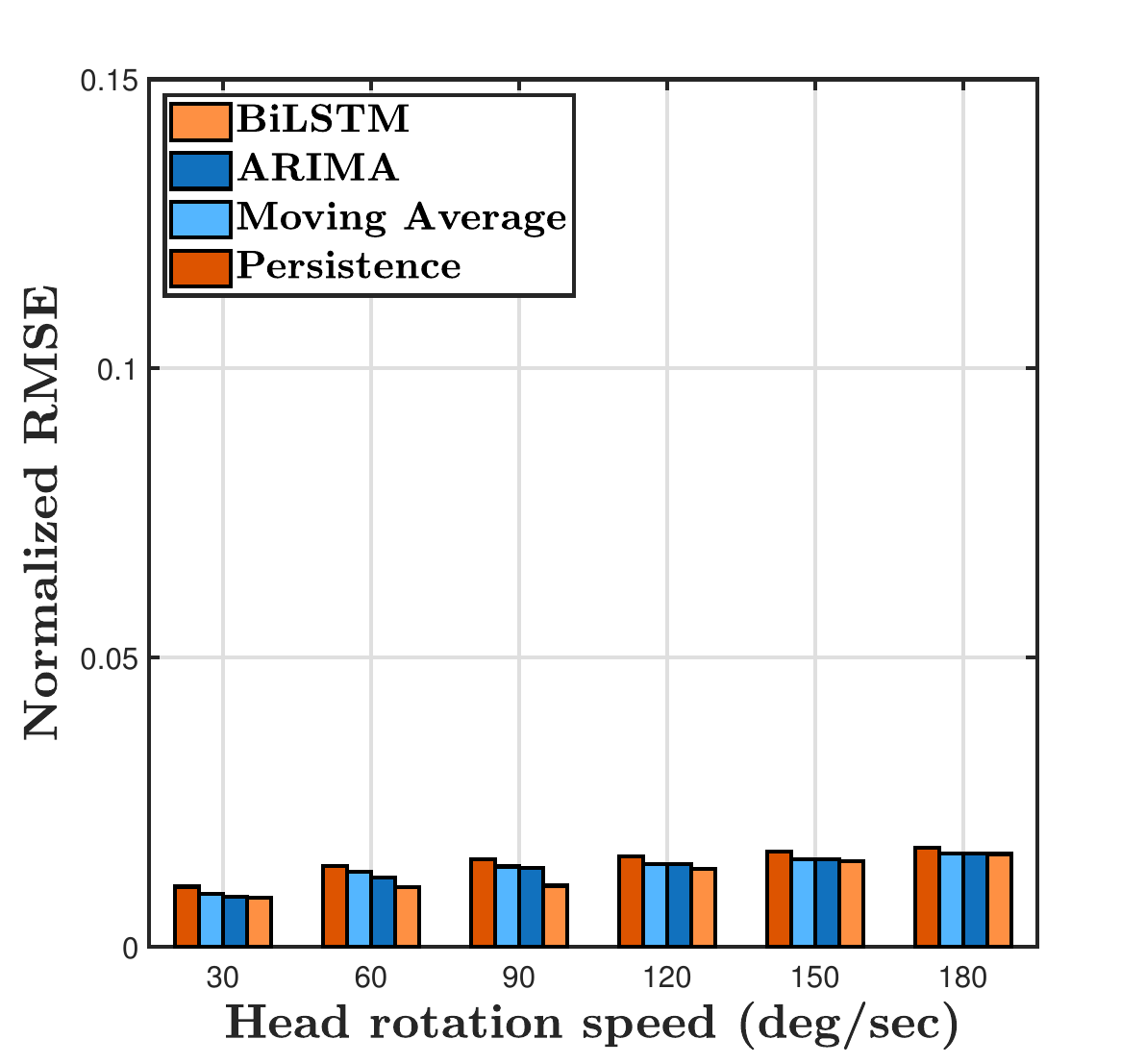}\label{rmmse_pitch6}%
  }
  \subfloat[Prediction of roll]{%
    \includegraphics[width=0.333\textwidth]{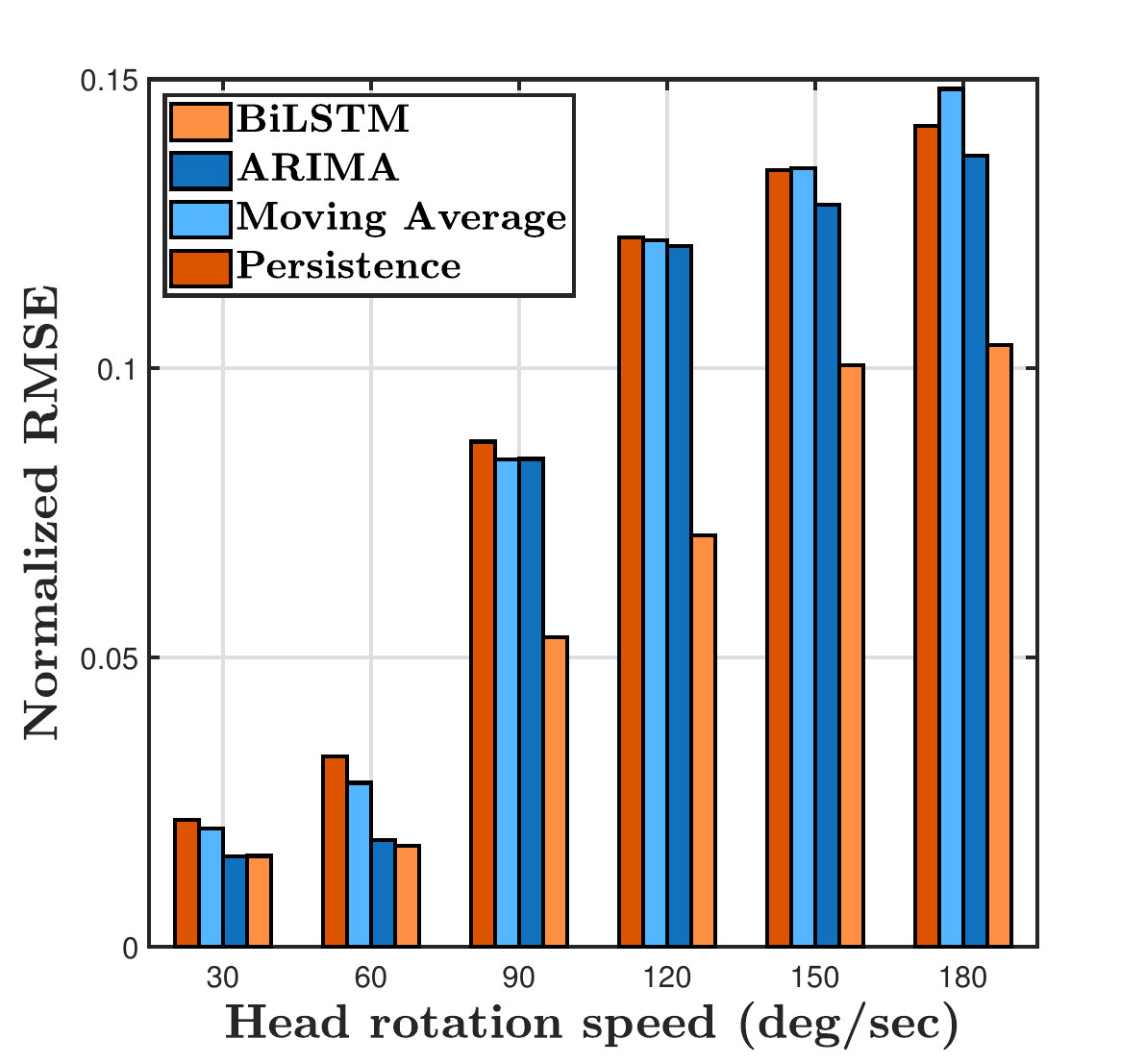}\label{rmmse_roll6}%
  }

  \caption{Comparison of normalized RMSE of different methods for a prediction horizon of 90 msec: (a) yaw, (b) pitch, and (c) roll.}
  \label{rmse_all6}
\end{figure*}
\setlength{\textfloatsep}{1pt}

\begin{figure*}[!t]
  \centering
  \subfloat[Prediction of yaw]{%
    \includegraphics[width=0.333\textwidth]{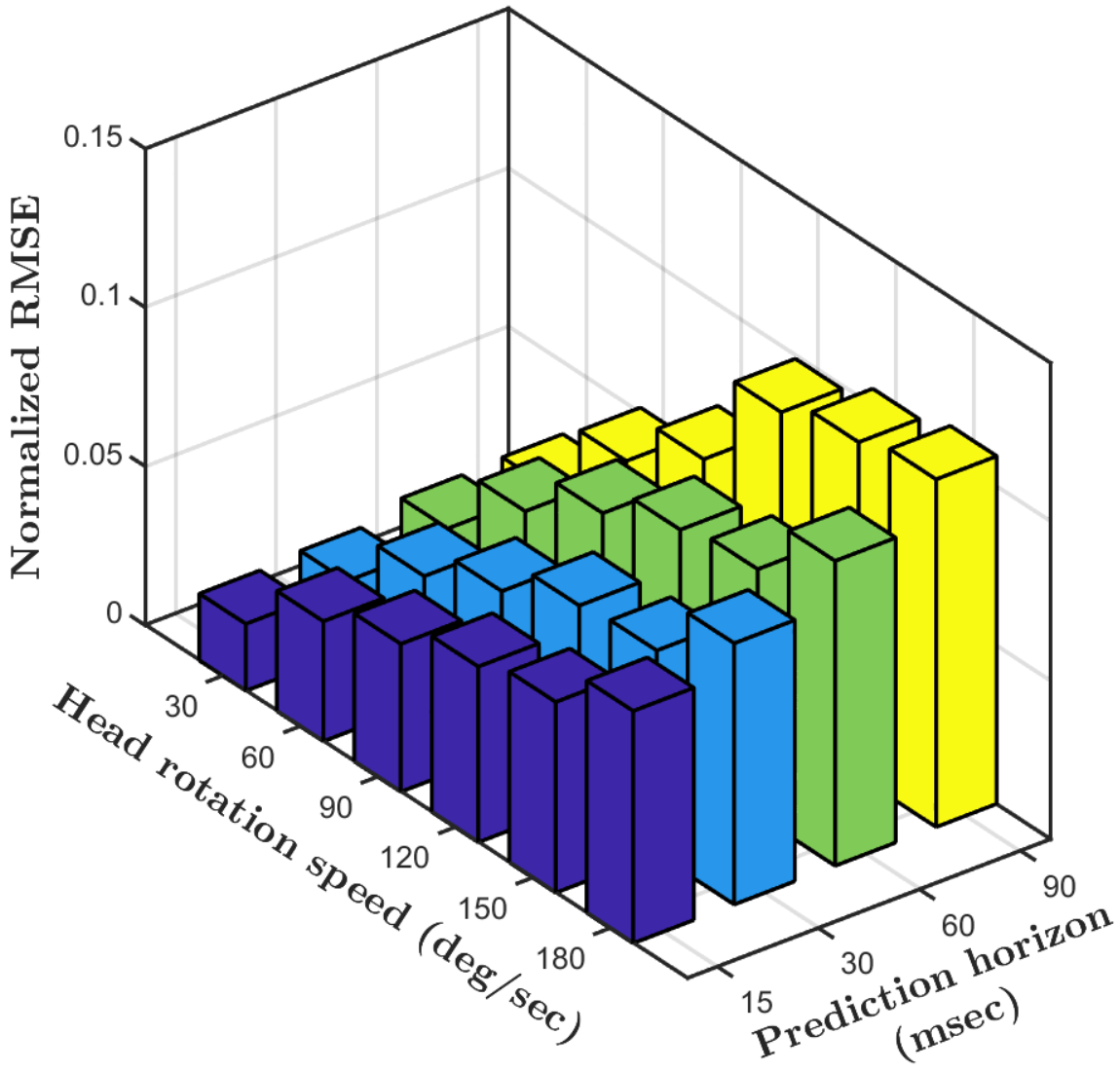}\label{trend1}%
  }
  \subfloat[Prediction of pitch]{%
    \includegraphics[width=0.333\textwidth]{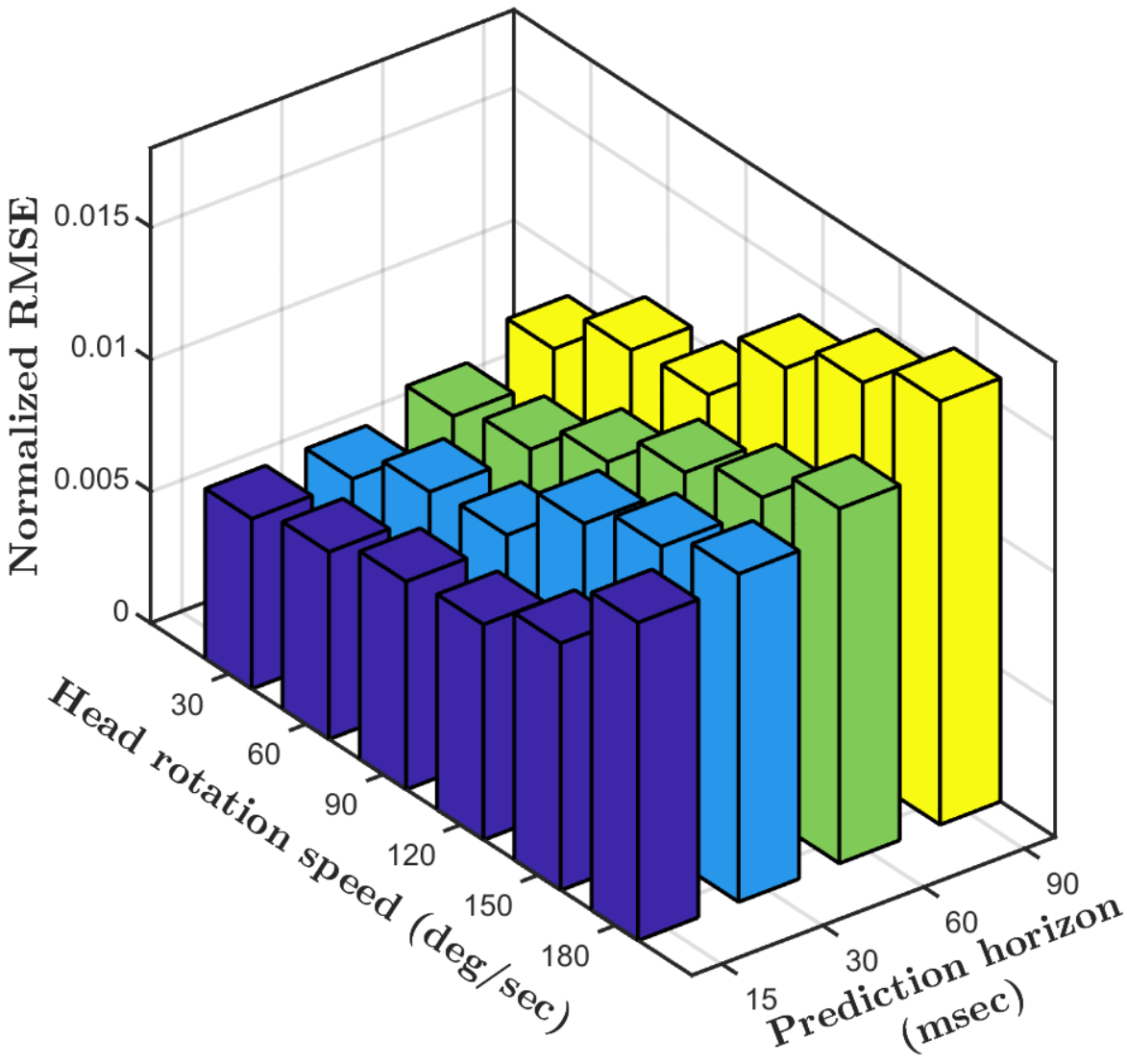}\label{trend2}%
  }
  \subfloat[Prediction of roll]{%
    \includegraphics[width=0.333\textwidth]{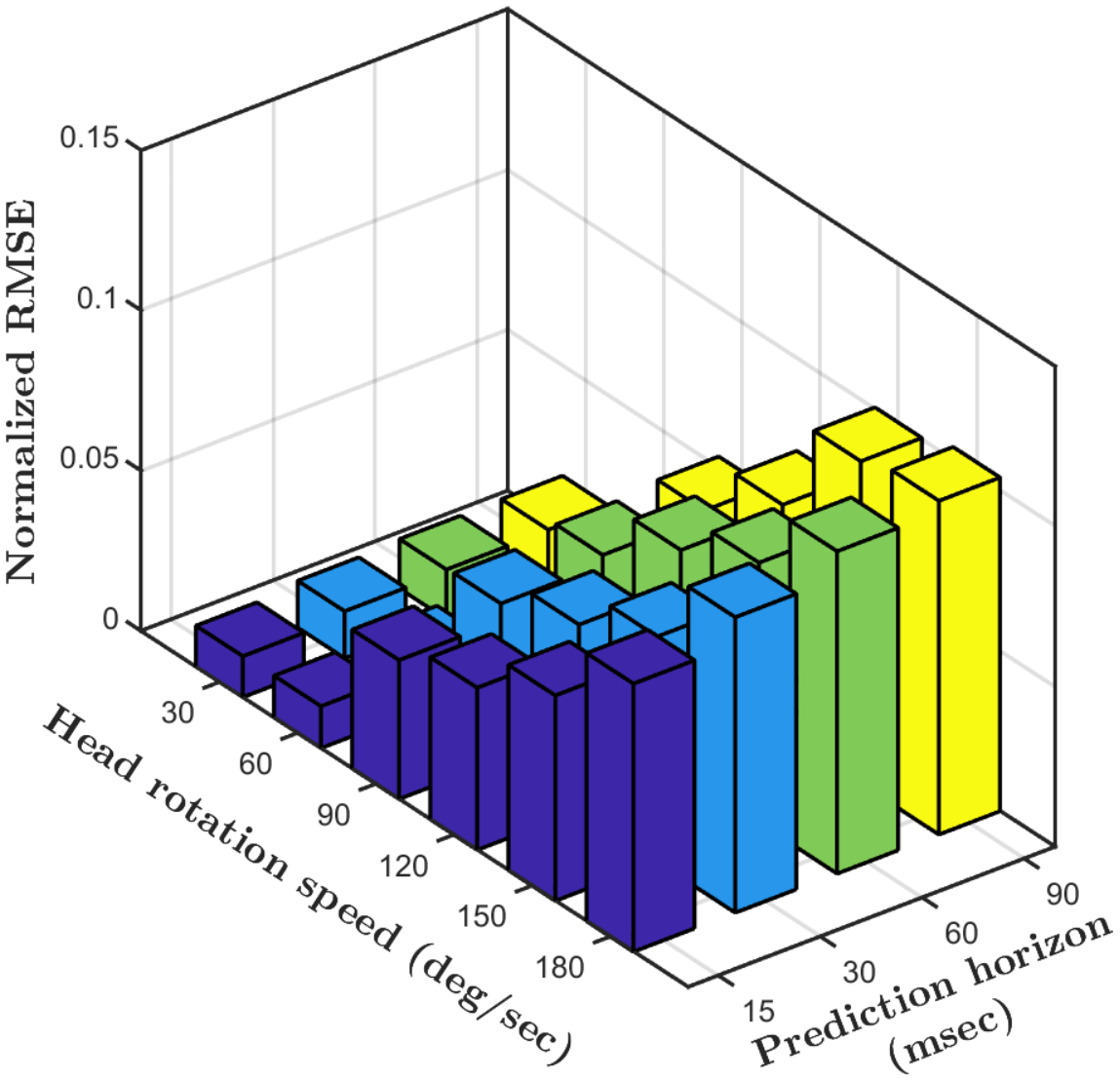}\label{trend3}%
  }

  \caption{Variation of normalized RMSE against average angular speed of human's head rotation and prediction horizon (msec ahead) using BiLSTM method on (a) yaw, (b) pitch, and (c) roll data streams.}
  \label{lstm_trends}
\end{figure*}
\setlength{\textfloatsep}{1pt}
\section{Performance Evaluation} \label{sec5}
In this section, firstly, we make a comparative analysis of the performance of different prediction methods implemented in MATLAB with HMD orientation data. Secondly, we discuss various insights obtained from discrete event-driven simulation of the proposed HMC-DBA scheme in OMNeT++.
\vspace{-\baselineskip}
\begin{figure*}[!t]
  \centering
  \subfloat[2K XR frame quality]{%
    \includegraphics[width=0.333\textwidth]{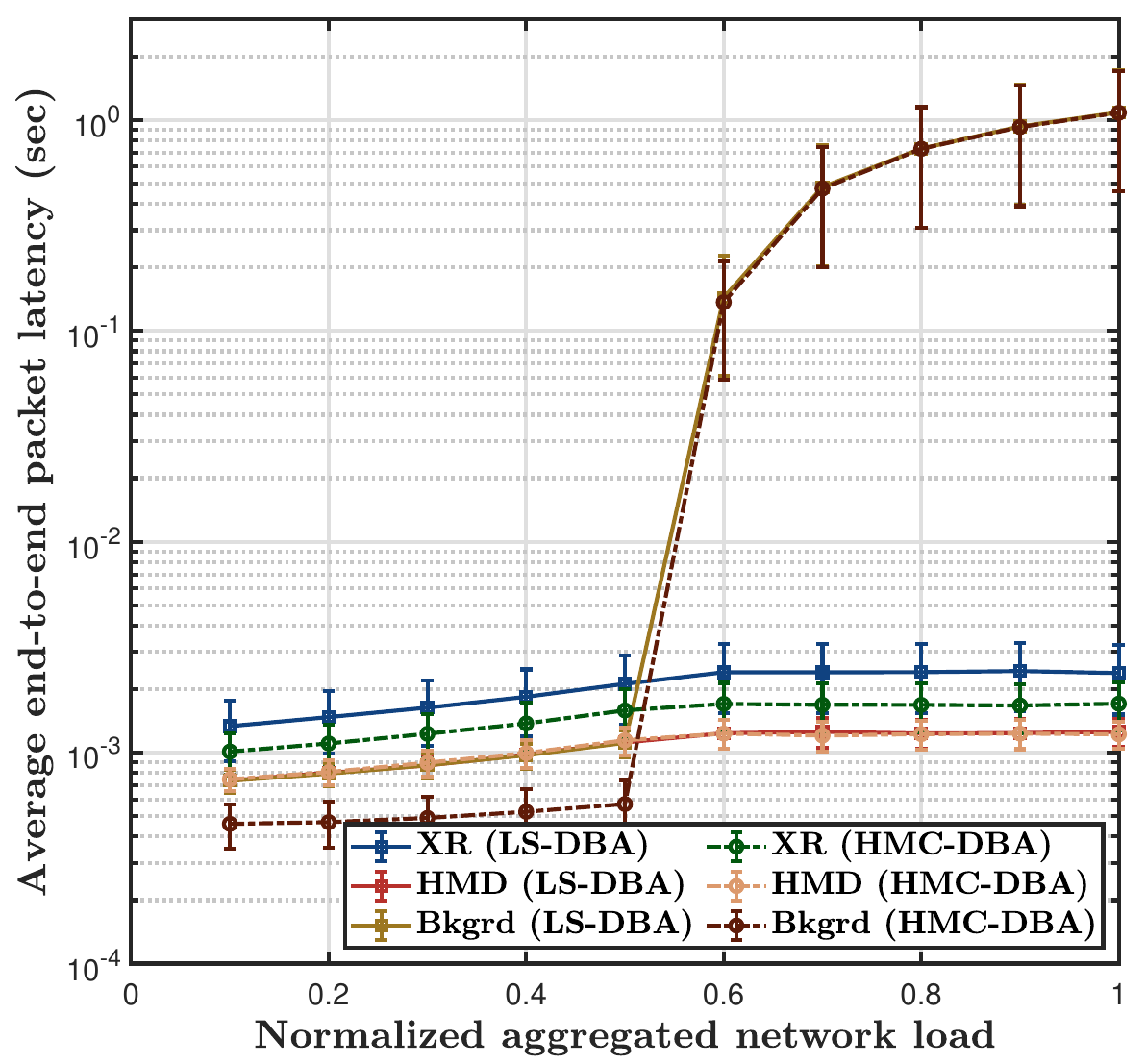}\label{e2e_2K_16}%
  }
  \subfloat[4K XR frame quality]{%
    \includegraphics[width=0.333\textwidth]{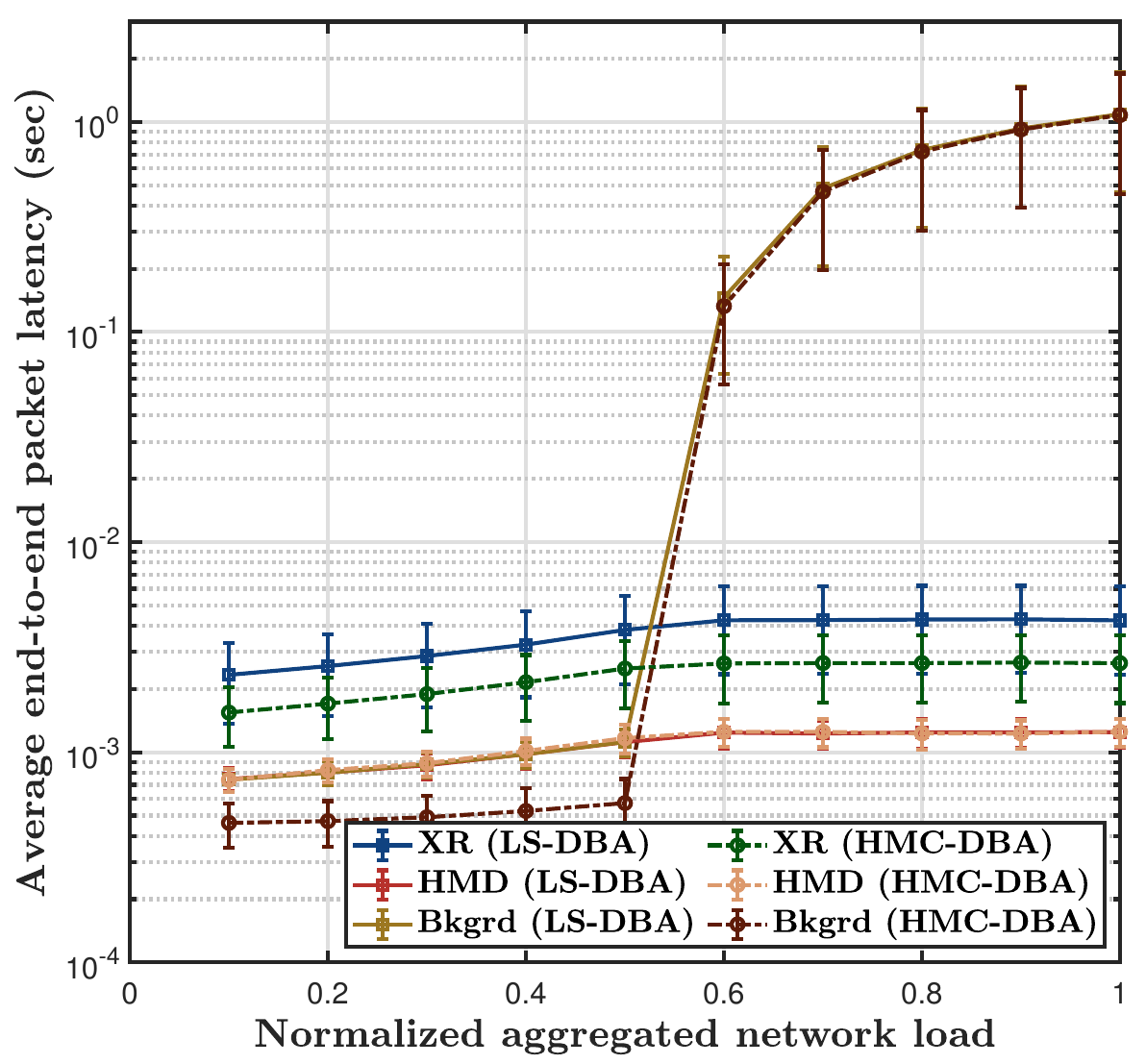}\label{e2e_4K_16}%
  }
  \subfloat[8K XR frame quality]{%
    \includegraphics[width=0.333\textwidth]{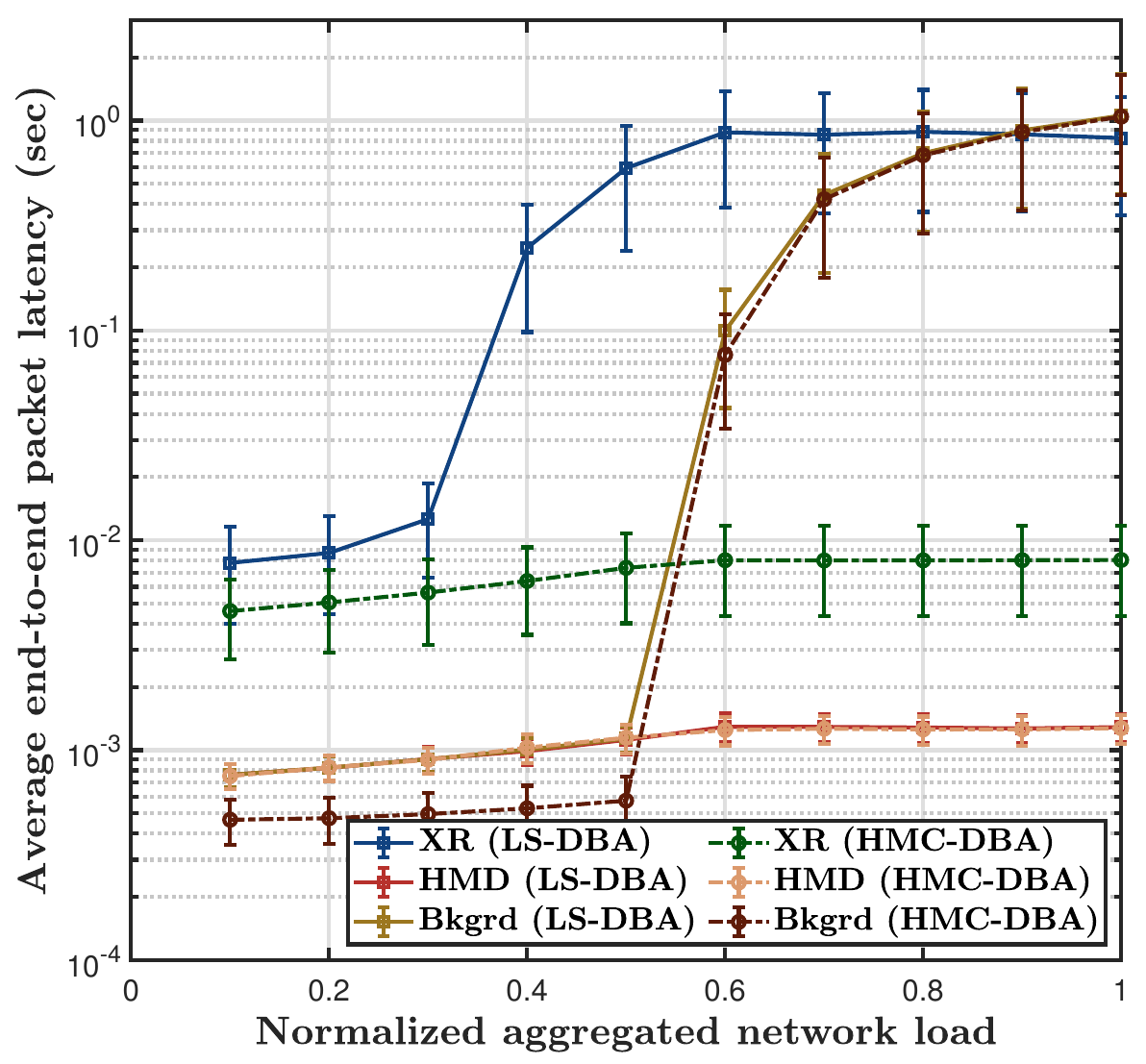}\label{e2e_8K_16}%
  }

  \caption{Comparison of end-to-end latency of XR, HMD, and background traffic packets with HMC-DBA against state-of-the-art baseline LS-DBA scheme over the FTTR-Business network architecture with FTTP (50G-EPON, 1:16) and FTTR (10G-EPON, 1:8) while transmitting XR frames of quality (a) 2K, (b) 4K, and (c) 8K.}
  \label{e2e_all_16}
\end{figure*}
\setlength{\textfloatsep}{1pt}

\begin{figure*}[!t]
  \centering
  \subfloat[2K XR frame quality]{%
    \includegraphics[width=0.333\textwidth]{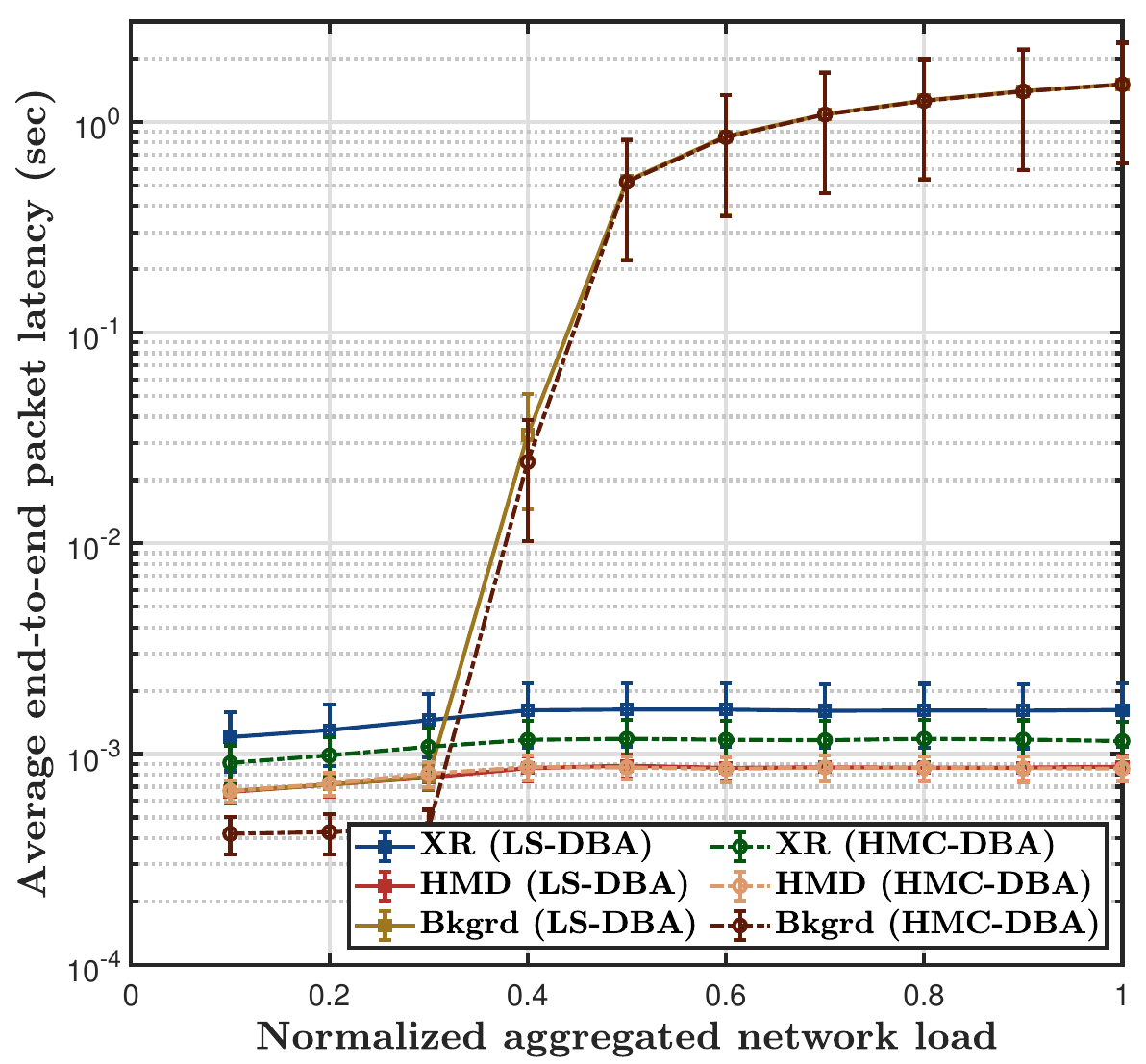}\label{e2e_2K_8}%
  }
  \subfloat[4K XR frame quality]{%
    \includegraphics[width=0.333\textwidth]{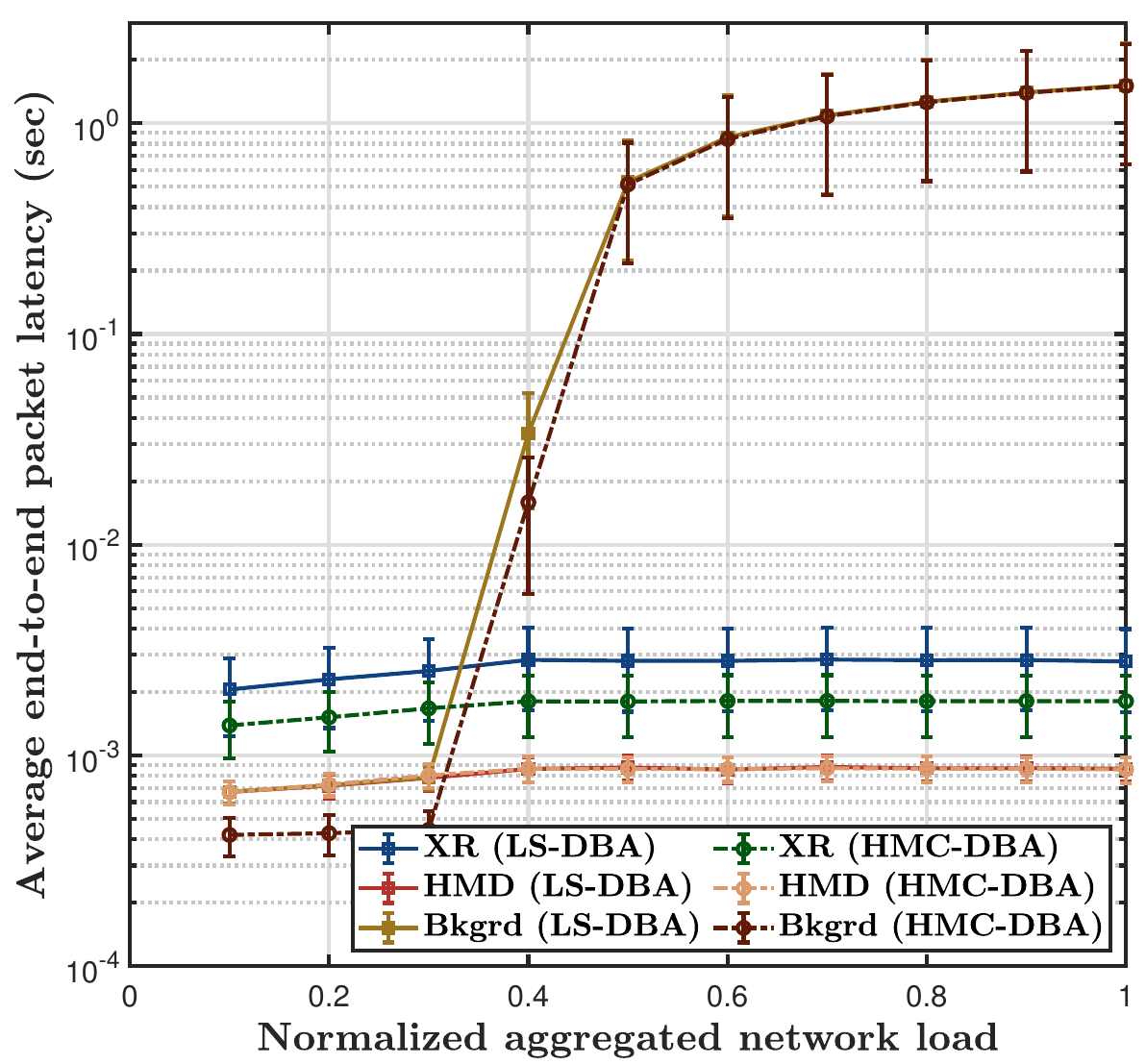}\label{e2e_4K_8}%
  }
  \subfloat[8K XR frame quality]{%
    \includegraphics[width=0.333\textwidth]{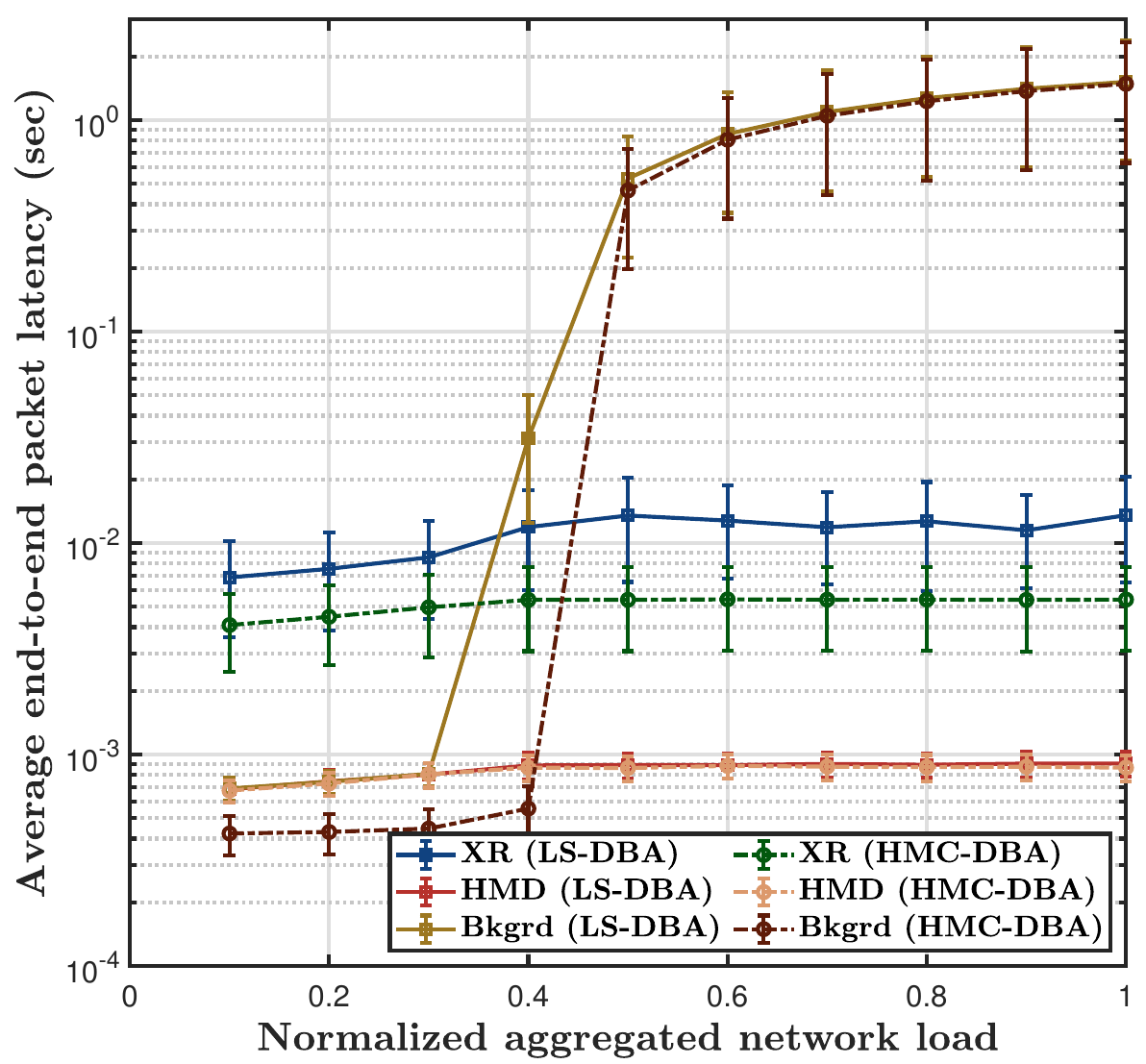}\label{e2e_8K_8}%
  }

  \caption{Comparison of end-to-end latency of XR, HMD, and background traffic packets with HMC-DBA against state-of-the-art baseline LS-DBA scheme over the FTTR-Business network architecture with FTTP (50G-EPON, 1:8) and FTTR (10G-EPON, 1:8) while transmitting XR frames of quality (a) 2K, (b) 4K, and (c) 8K.}
  \label{e2e_all_8}
\end{figure*}
\setlength{\textfloatsep}{1pt}
\subsection{HMD Dataset and Network Simulation Configurations} \label{sec5.1}
\textcolor{black}{From our lab experiments, as described in Sec.~\ref{sec3.2}, we collected approximately $10^6$ human head movement data samples (yaw, pitch, roll) while performing typical actions during a H2M collaboration like picking up some object, placing an object on a rack, tightening a screw with a screwdriver, to name a few \cite{ind_act}. The primary characteristics of this data is that the human head movements do not happen continuously but intermittently like a random on-off process. Hence, instead of analyzing the dataset based on performed actions, we classify and divide this entire dataset according to the average speed of head movement following (\ref{eq01}), e.g., 30, 60, 90, 120, 150, 180 degree/sec. This approach provides us with a generalized framework that can be used to train and test the considered prediction models on datasets having similar characteristics. We considered the open-source data shared by the authors of \cite{head_data}, consisting of head movement tracking information of 48 participants immersively watching 18 different sphere videos. The size of this database is approximately 228 MB, consisting of approximately $30\times10^6$ samples.}\par
\vspace{-0.3\baselineskip}
\textcolor{black}{For the simulation of HMC-DBA scheme, following the ITU-T recommendations \cite{itu_fttr}, we consider a 50G-EPON of 20 km length as the first-segment FTTP network with the maximum available datarate in both the uplink and downlink is 50 Gbps. The duration of the maximum polling cycle is 0.5 msec and the guard time between packets from consecutive MFUs is 1 $\mu$sec. The number of MFUs supported are 8 (split-ratio 1:8) and 16 (split-ratio 1:16). Each of these MFUs control one FTTR segment with 10G-EPON with maximum uplink and downlink datarate 10 Gbps, split ratio 1:8, MFU-SFU distance 20 meters, duration of the maximum polling cycle 2 msec, and the guard time between packets from consecutive SFUs is 2 $\mu$sec. This implies that there are maximum $(16\times8) = 128$ SFUs, each connected to a WiFi 6/6E WAP (maximum throughput 9.6 Gbps in in-premise environments \cite{WiFi6}) in this network setting. Furthermore, we consider that there are 56 pairs of human-machine collaborators ($56\times2 = 112$ in total) such that the remaining 16 SFUs support background traffic from several wireless devices. The XR traffic with 2K, 4K, and 8K frame qualities are generated following Table~\ref{table1}. The HMD traffic is generated following a Gamma distribution with Mean 15 msec, as detailed in Sec.~\ref{sec3.2}.}

\vspace{-\baselineskip}
\subsection{Results and Discussions} \label{sec5.2}
In Fig.~\ref{rmse_all6}, we show normalized RMSE of human's head movement prediction using persistence, moving average, ARIMA, and BiLSTM methods with a prediction horizon of 90 msec (6 samples ahead). We perform training and testing separately on the three components for head orientation tracking, i.e., yaw, pitch, and roll. Note that, for \emph{persistence}, the predicted value is the same as the last observed value; for \emph{moving average}, the predicted value is the average of samples observed in last 1 sec; for \emph{ARIMA}, training is done with a ARIMA(6,1,2) model on the samples arrived in first 5 min to dynamically predict the rest; and for \emph{BiLSTM}, training is done on nearly $10^6$ samples with a window of 6 past samples for each speed of head rotation and prediction horizon value. \textcolor{black}{Note that the traffic prediction models are trained offline based on historical data. The highest training time was consumed by the BiLSTM networks (around ~6-8 hours) for each prediction horizon on a laptop with Intel Core i7-1065G7 (Quad-core, 10th Gen) processor, 32 GB LPDDR4x RAM, and NVIDIA GeForce GTX 1650 GPU. However, the inference time on each sample was a few micro-seconds only. Therefore, with the support of practical Edge-AI servers, the effect of traffic prediction time will be negligible.}\par
\vspace{-0.5\baselineskip}
In Fig.~\ref{rmmse_yaw6}, we observe that the normalized RMSE using persistence is highest among all for yaw. However, as the speed of head rotation increases, the performance of moving average starts to degrade due to failure to track sudden fast swings. It is important to note that when the human's head movement is significantly faster than the camera motors, then longer prediction horizons are required with very high prediction accuracy. ARIMA has shown to have comparable performance at lower speeds of head rotation and slightly better performance than persistence and moving average at higher speeds. BiLSTM performs the best among all methods, achieving approximately 50\% lower normalized RMSE compared to persistence. Similar results are also observed with roll data, as shown in Fig.~\ref{rmmse_roll6}. For pitch data, normalized RMSE of all the methods are much lower (almost 10 times) compared to yaw and roll, as shown in Fig.~\ref{rmmse_pitch6}, because the swing is within 50\% range only.\par
When the human head movements are much faster than the machine's camera movements, then a suitable prediction horizon is required to compensate this gap. For example, if both the human's head and machine's camera rotate at 15 degree/sec, then at least 1 sample (average 15 msec, as per Sec.~\ref{sec3.2}) ahead prediction is required while the network ensuring an end-to-end packet latency budget of 10-15 msec \cite{xr_std}. However, if the machine's camera rotates at 15 degree/sec but the human's head rotates at 30 degree/sec, 60 degree/sec, and 90 degree/sec, then at least 2 samples (average 30 msec) ahead, 4 samples (average 60 msec) ahead, and 6 samples (average 90 msec) ahead predictions are required, respectively. \textcolor{black}{Therefore, in Fig.~\ref{lstm_trends}, to provide a sensitivity analysis on the choice of prediction horizons, we show the variation of normalized RMSE of BiLSTM method with yaw, pitch, and roll data streams against head rotation speed and prediction horizon.} We observe an increasing trend of normalized RMSE as both these variables increase but at different rates.\par
Next, in Fig.~\ref{e2e_all_16}, we compare average end-to-end latency of packets from XR frames, HMD orientations, and background traffic over the FTTR-Business network with FTTP (50G-EPON, 1:16) and FTTR (10G-EPON, 1:8). In Figs.~\ref{e2e_2K_16}, \ref{e2e_4K_16}, \ref{e2e_8K_16}, we observe that the end-to-end latency requirements for \emph{fair QoE} and \emph{comfortable QoE} (refer to Table~\ref{table1}) are satisfied with 2K and 4K quality XR frames, respectively, but is completely violated for \emph{ideal QoE} with 8K quality XR frames by LS-DBA. Moreover, the jitter standard deviation for 2K and 4K quality XR frames is $\sim$2 msec, but higher for 8K quality XR. On the other hand, our proposed HMC-DBA ensures a much lower end-to-end latency with 2K and 4K quality XR frames as well as satisfies the \emph{ideal QoE} requirements with 8K quality XR frames (borderline). The average end-to-end latency of HMD orientation data in all these plots are well within the QoE requirements due to its very low packet size. Nonetheless, the end-to-end latency of background traffic is relatively lower than HMD and XR traffic at low and medium load conditions, but shoots up around 50\% aggregated network load. Based on these observations, in Fig.~\ref{e2e_all_8}, we compare average end-to-end latency of packets from XR frames, HMD orientations, and background traffic over FTTR-Business network with FTTP (50G-EPON, 1:8) and FTTR (10G-EPON, 1:8). With this network configuration, we make similar observations for 2K and 4K quality XR frames. Although, the end-to-end packet latency for 8K quality XR frames is slightly higher than the ideal QoE requirements with LS-DBA scheme, our proposed HMC-DBA is able to ensure the ideal QoE requirements successfully.\par
%
\begin{figure}[!t]
\centering
\includegraphics[width=0.8\columnwidth,keepaspectratio]{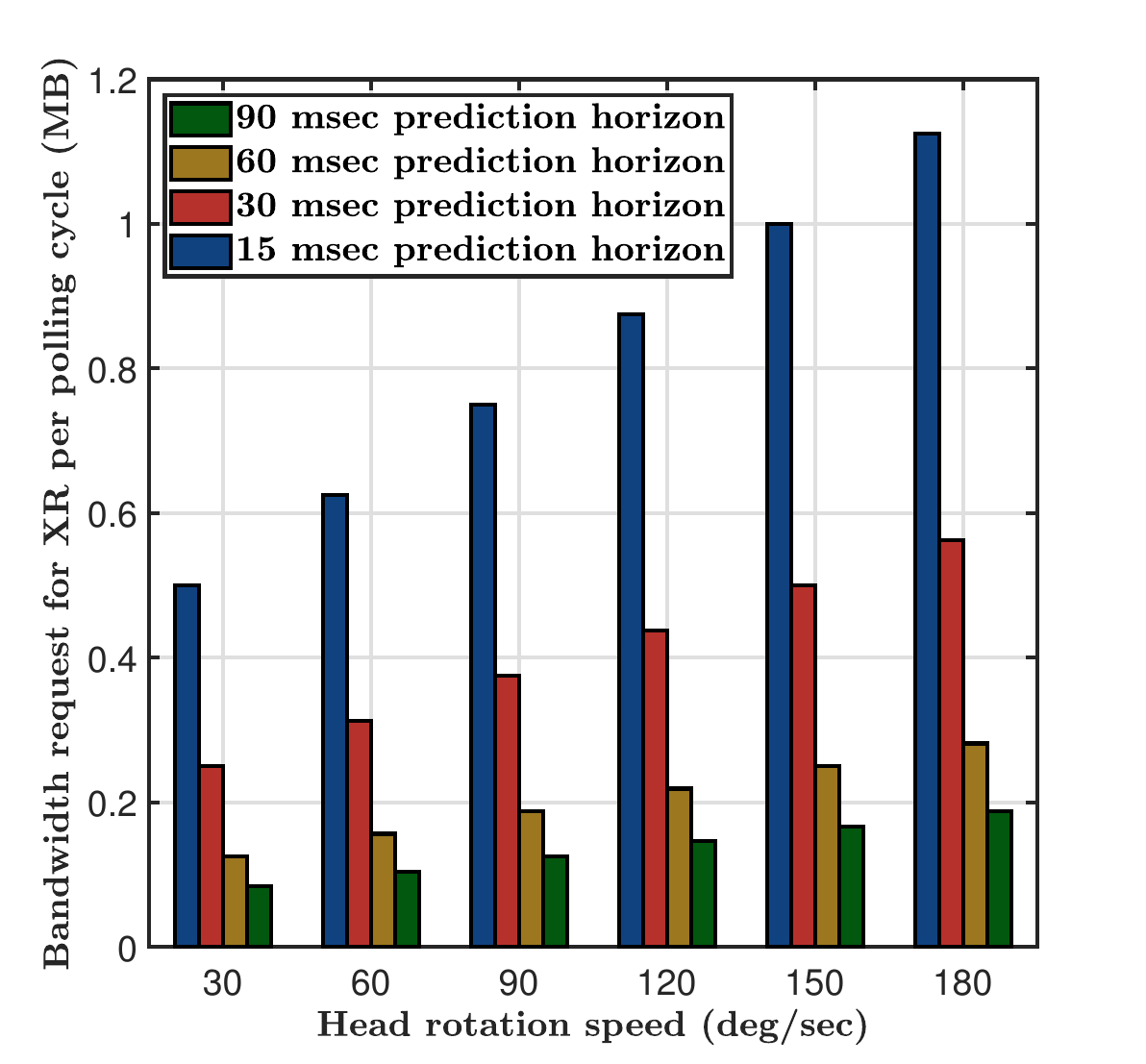}
\caption{Reduced bandwidth request from SFUs for transmitting XR frames of 8K quality by predicting future human head movements.}
\label{benefit}
\end{figure}
\setlength{\textfloatsep}{5pt}
The fundamental reason for lower latency and jitter with HMC-DBA in the transmission of XR frames, is reduced bandwidth request for transmitting each XR frame. In Fig.~\ref{benefit}, we show the variation of bandwidth request per 8K quality XR frame against head movement speed. When head rotation speed increases, then consecutive XR frame size also increases. If direct synchronization between the human's head and machine's camera was possible, then the bandwidth requests from SFUs to transmit XR frames would also increase. However, prediction of head movements before the camera to be reoriented to the same angular position over a longer time duration, according to the chosen prediction horizon. This also helps to compensate for the slower camera movement speed. For a 15 msec prediction horizon, there is no gain as the entire frame needs to be transmitted according to the head movement. However, for prediction horizons of 30 msec, 60 msec, and 90 msec, significant reduction of frame sizes is possible.

\section{Conclusion} \label{sec6}
In this paper, a novel HMC-DBA scheduling scheme for industrial H2M collaborations over FTTR-Business networks has been proposed. A normalized RMSE of less than 0.1 has been achieved for prediction of all the Euler angle components (yaw, pitch, and roll) of human's head movements using the BiLSTM method. Thorough numerical evaluations have been performed to comparatively establish that BiLSTM is the best predictor of the Euler angle components of human's head movements over persistence, moving average, and ARIMA. Moreover, using prediction of human's future head movements and the gap of rotation speed between human's head and machine's camera, XR frames of much smaller size could be generated during fast head movements with high angular shifts. This ensured the low-latency and low-jitter requirements for XR frame transmission with fair, ideal, and comfortable QoE. In addition, the time instants of XR and HMD data generation have been estimated exploiting their pseudo-periodic properties. By incorporating the estimated XR frame sizes and arrival time instants in our proposed HMC-DBA scheme, the ideal QoE of immersive experience for industrial H2M collaborations over FTTR-Business networks has been ensured, which was impossible with state-of-the-art baseline bandwidth allocation schemes.

\bibliographystyle{IEEEtran}
\bibliography{IEEEabrv,references}
\begin{IEEEbiographynophoto}{Sourav Mondal}
(GS'16–M'21) received PhD from the University of Melbourne in 2020, M.Tech from Indian Institute of Technology Kharagpur in 2014, and B.Tech from Kalyani Govt. Engineering College, affiliated to West Bengal University of Technology in 2012. Currently, he is working as a Research Fellow in the Faculty of Engineering and IT, Department of Electrical and Electronic Engineering of the University of Melbourne. Prior to this, he briefly worked as an EDGE/Marie Skłodowska-Curie Post-doctoral Fellow in Trinity College Dublin, Ireland and as a Senior Lead Systems Engineer (Corporate R\&D) in Qualcomm Bengaluru Design Centre, India.
\end{IEEEbiographynophoto} 


\begin{IEEEbiographynophoto}{Elaine Wong}
(M'02–SM'14) received B.E. and Ph.D. degrees from the University of Melbourne, Australia. She is currently Associate Dean and Professor at the Faculty of Engineering and Information Technology, University of Melbourne. Her research interests include energy-efficient optical and wireless networks, optical-wireless integration, broadband applications of vertical-cavity surface-emitting lasers, wireless sensor body area networks, and emerging optical and wireless technologies for human-to-machine applications. She has coauthored more than 150 journal and conference publications. She has served on the editorial board of the \emph{Journal of Lightwave Technology} and the \emph{Journal of Optical Communications and Networking}.
\end{IEEEbiographynophoto}

\vfill

\end{document}